\documentclass{aa}
\usepackage{graphics}
\usepackage{epsfig}

\def\msol{M_\odot}
\def\mmin{m_{\rm min}}
\def\mv{M_{\rm V}}
\def\te{T_{\rm eff}}
\def\lsol{L_\odot}
\def\simgr{\,\hbox{\hbox{$ > $}\kern -0.8em \lower 1.0ex\hbox{$\sim$}}\,}
\def\simle{\,\hbox{\hbox{$ < $}\kern -0.8em \lower 1.0ex\hbox{$\sim$}}\,}
\def\beq{\begin{equation}}
\def\eeq{\end{equation}}

\def\apj{ApJ}                 
\def\apjs{ApJS}               
\def\aap{A\&A}                
\def\mnras{MNRAS}             

\begin{document}

\thesaurus{08.05.03, 08.22.1}

\title{Period - luminosity - color - radius relationships
 of Cepheids as a function of metallicity: evolutionary effects.}

\author{Yann Alibert\inst{1} \and Isabelle Baraffe\inst{1,2} \and Peter 
Hauschildt\inst{3} \and France Allard\inst{1}}

\offprints{Y. Alibert}

\institute{C.R.A.L (UMR 5574 CNRS), 
 Ecole Normale Sup\'erieure, 69364 Lyon
Cedex 07, France\\ email: yalibert, ibaraffe, fallard @ens-lyon.fr
\and Max-Planck Institut f\"ur Astrophysik, Karl-Schwarzschildstr.1,
D-85748 Garching, Germany
 \and
Dept. of Physics and Astronomy and Center for Simulational Physics,
University of Georgia Athens, GA 30602-2451 \\ email:
yeti@hobbes.physast.uga.edu}

\date{Received /Accepted}

\titlerunning{Period-luminosity-color-radius relationships of Cepheids}
\authorrunning{Alibert et al. } 
\maketitle

\begin{abstract}

Based on consistent evolutionary and pulsation calculations,
we analyse the effect of metallicity and of  different convection
treatments in the stellar models on period - magnitude, - color
and - radius relationships. In order to perform an accurate comparison with observations, we have  computed
grids of atmosphere models and synthetic spectra for
 different metallicities,
 covering the range of effective temperatures
and gravities relevant for Cepheids. 
The models are compared to recent observations of galactic
and Magellanic Clouds Cepheids. Unprecedented level of agreement is
found between models and observations.

We show that within the range of metallicity
for the Galaxy and the Magellanic Clouds, a change of slope
in the period - luminosity (PL) relationship
is predicted at low periods, due to the reduction of
 the blue loop  during core He burning. 
The minimum mass undergoing a blue loop and consequently
the critical period at
which this change of slope occurs depend on the metallicity Z
and on the convection treatment in the stellar models.

However, besides this change
of slope, we do not
find any significant effect of metallicity on
period - magnitude relationships from V to K bands,
and on period  - color relationships in IR colors.
We only find a detectable effect of $Z$ 
on $(B-V)$ colors.
These results are not
affected by uncertainties inherent to current stellar models,
mainly due to convection treatment.
\end{abstract}

\section{Introduction}

Although Cepheids have been studied extensively since the discovery by
Leavitt in 1908 of their fundamental period -
luminosity (PL) relationship, the
metallicity dependence of this relation 
still remains unsettled (see Tanvir 1997 for a
review). 
Since the Cepheid PL relationship is a cornerstone to extragalactic distance
scales, the existence of such dependence is of fundamental
importance.

This old debate has recently been revived due to high-quality
observations
from the HST, Hipparcos and
the micro-lensing experiments which collected large samples of Cepheid
data in the Magellanic Clouds (EROS: Renault et al.
1996, Beaulieu \& Sasselov 1996 and references therein;
 MACHO: Welch et al. 1996 and
references therein). The forthcoming release of the OGLE-2 data
(Udalski, Kubiak, Szymanski 1997 and references therein) will
bring further important elements to this debate.
Based on the EROS Cepheid sample, a comparison of the observed PL relations
in the Small (SMC) and the Large (LMC) Magellanic Clouds
led Sasselov et al. (1997) to suggest  a significant metallicity
effect.
In the same vein, Sekiguchi and Fukigita (1998) reanalysed Cepheid
data used as calibrators and claimed also a large metallicity
effect. 
Such controversy with previous claims (see Tanvir 1997)
 illustrates the difficulty to disentangle
a metallicity effect from other effects such 
as reddening or distance uncertainties and to reach definitive conclusions. 

Since empirical calibrations do not provide a clear answer to
the metallicity correction, a theoretical analysis could
provide the basis for a better understanding of this problem.
However, uncertainties inherent to stellar
evolution models (convection treatment, overshooting, opacities, etc..),
pulsation calculations (nonlinear effects, time-dependent convection)
and atmosphere models required to determine the bolometric corrections
render the task difficult.

One of the most systematic study 
has been 
performed by Chiosi, Wood \& Capitanio (1993), based on linear
stability analysis of a large grid of models with different
compositions, effective temperatures, masses and mass - luminosity (ML)
relationships. The metallicity effect on the PL
relationship was found to be rather small.
However, the models of Chiosi et al.
(1993) are based on the old Los Alamos opacity Library (Huebner et al.
1977), which is a key ingredient to the stellar
and pulsation models. Since then, opacity calculations have greatly
improved, thanks to both the
Livermore group (OPAL,
Iglesias \& Rogers 1991;  Iglesias, Rogers, \& Wilson 1992) and the OP
project (Seaton et al. 1994).
More recently, Iglesias and Rogers (1996) have updated the OPAL
opacities, with an increase of the opacity up to
20\% for solar-type mixtures. Adopting these improved opacities, 
Saio and Gautschy (1998) derived  theoretical PL relationships
based on consistent stellar evolution and linear pulsation calculations
for different metallicity and masses ranging from 4 to 10 $\msol$.
They find  a negligible metallicity
dependence of the PL relationship.
On the other hand, Bono et al. (1998a, BCCM98) recently investigated the PL and
period - color (PC) relationships on the basis of non-linear
calculations.
Their analysis, which covers a mass range from 5 to 11 $\msol$  
predicts a significant dependence of the PL relationship on
metals, in contradiction with the results of Saio and Gautschy (1998), although
for similar mass and metallicity ranges.   

In parallel with the work of Saio and Gautschy (1998), Baraffe et al. (1998)
performed self-consistent calculations between stellar
evolution and linear stability analysis for the SMC, the LMC and the Galaxy.
This work was mainly devoted to the comparison with the observed
PL relationships and the Beat Cepheids of the SMC and LMC data
obtained by  the EROS and MACHO groups. 
Baraffe et al. (1998) have emphasized the importance of the first
crossing
instability phase, which corresponds to models which are just leaving
the Main Sequence and evolving toward the Red Giant Branch on
a thermal time-scale. Such models can explain most of the
first-overtone/second-overtone (1H/2H) beat Cepheids observed in the Magellanic
Clouds (MC).
An other important result stressed in 
Baraffe et al. (1998) is the possible explanation by pure evolutionary
effects of the 
change of slope in the PL relationship
 observed  by Bauer et al. (1998)
for short period SMC Cepheids. Such results show the necessity to take
into account evolutionary properties in order to derive reliable PL
relationships and reproduce observations. 

The goal of the present paper is to analyse carefully the PL
relationships based on the evolutionary models of Baraffe et
al. (1998), taking into account the effect of chemical composition
and of different convection treatments on the evolutionary models (\S 2).
Baraffe et al. (1998) did not perform such detailed analysis, since 
their bolometric corrections were derived from atmosphere models
with constant gravity.
For the purpose of the present work, we have calculated a grid
of atmosphere models covering the required range of effective
temperatures and gravities and  for the same metallicities as  in
the evolutionary models. 
This enables us
to analyse fine effects of metallicity on the derived synthetic
colors. In \S 3, we derive theoretical
PL relationships and \S 4 is devoted to the comparison with observed 
period-magnitude-color-radius relationships in the SMC, LMC and the Galaxy. 
We construct also theoretical period histograms
and compare them
to the observed histograms derived from  sample of Cepheids in
the MC's (cf. \S4).
Finally, section 5 is devoted to the analysis of
metallicity effects and to discussion and \S 6 to conclusions.

\section {Evolutionary calculations}

Stellar models are computed with the Lyon evolutionary Henyey-type
code,  
originally developed at the G\"ottingen Observatory (Baraffe and El
Eid 1991).  
The most recent OPAL opacities (Iglesias \& Rogers 1996) are used for the
inner structure
$T>6000 K$. For temperatures lower than $6000K$, we use the Alexander
\&  Fergusson (1994) opacities.
Mass loss is taken into account according to de Jager et al. (1988) 
with a scaling factor which depends on
the metallicity as ($Z/Z_{\odot})^{0.5}$, as indicated by stellar wind
models (Kudritzki et al. 1987, 1991). The effect of mass loss is
however negligible during the evolutionary phases involved, and in any
case the total mass lost does not exceed 2 \% of the initial mass. 

Convective transport is computed using the mixing-length theory
formalism as given by Kippenhahn \& Weigert (1990).  The ratio
$\alpha_{\rm mix}$ 
of the mixing length ($l_{\rm mix}$) to the pressure scale height
($H_{\rm P}$) is
 fixed to 1.5. 
In order to determine the uncertainties resulting from a
variation of $l_{\rm mix}$, we have also 
computed two sets of evolutionary tracks with $\alpha_{\rm mix} = 2$
for metallicities representative
of the MC's, for which more Cepheid data (fundamental
and overtone pulsators) are available than for the Galaxy.

The onset of convective instability is based on the Schwarzschild
criterion. Our main grid of evolutionary models is computed
with standard input physics (e.g without overshooting,  semiconvection,
or rotation induced mixing). In order to test the effect of some enhanced mixing
in the convective core, as expected when using overshooting,
we have also computed models in which the size of the convective core is
arbitrarily increased by an amount $d_{\rm ov} = \alpha_{\rm ov} H_{\rm P}$, where
 $\alpha_{\rm ov}$ is 
a free parameter fixed to 0.15, as justified in \S 2.3 and \S 4.5. 
This extended region is assumed to be completely mixed, but the
temperature gradient is supposed to be radiative.
This type of simple
prescription
is usually used to mimic the effect of overshooting or any other
source of enhanced  mixing beyond the standard convective
boundary (cf. El Eid, 1994; Pols et al. 1998).

We have computed stellar models in the mass range $2.75 - 12 M_{\odot}$ with
various chemical compositions $(Z, Y) = (0.02, 0.28)$, $(0.01, 0.25)$
and  $(0.004, 0.25)$ \footnote {$Z$ is the metal mass fraction 
 and $Y$ the helium mass fraction}, representative of respectively the Galactic,
LMC and SMC environments. 
In the case of the LMC, we have also computed different grids of
models with various  $Z$ and $Y$, 
{ \it i.e. } $(Z,Y)=(0.01,0.25)$, $(Z,Y)=(0.01,0.28)$, and
$(Z,Y)=(0.008,0.25)$.

\subsection{Effect of the initial chemical composition}

Since the present paper is devoted to period - luminosity relationships rather than
 to stellar evolutionary models, we concentrate in the following
 on properties relevant to Cepheids.
The first important effect of metallicity concerns the determination 
of the minimum mass $m_{min}$ which undergoes a blue loop entering
the instability strip (IS). 
As shown in
Fig. 1, $\mmin$ decreases with metallicity. We find for the fundamental
mode IS
$\mmin ~\sim 3 \msol$ for $Z$=0.004, $\mmin ~\sim 3.875 \msol$ for $Z$=0.01
 and
$\mmin ~\sim 4.75 \msol$ for $Z$=0.02. Note that Baraffe et al.
(1998) mention slightly lower $\mmin$ for $Z$=0.01 and $Z$=0.02,
but as shown in \S 3, these lower values are too uncertain due
to the arbitrary criterion used to determine the red edge.
Fig. 1 displays the evolutionary tracks of 
stars of mass $\mmin$ for different metallicities and the location
of fundamental unstable models (see \S 3 for the description
of the linear stability analysis). Stars with
masses just below
 $\mmin$ undergo a reduced blue loop which does not enter the instability
 strip. Note that all stellar models in Fig. 1 show
fundamental unstable modes during the first crossing of
the Hertzsprung-Russel diagram (HRD) toward the  Red Giant Branch
 (see Baraffe et al. 1998 for details).

\begin{figure}
\psfig{file=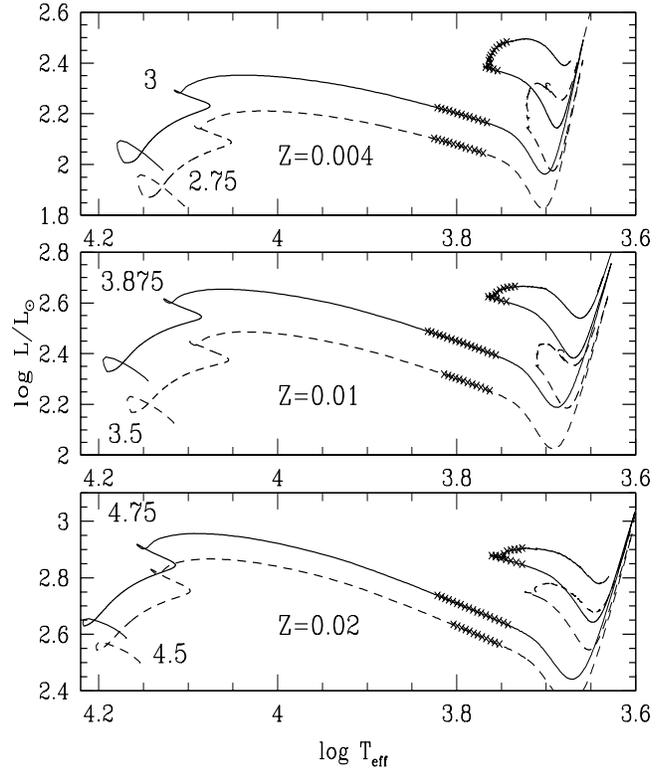,height=110mm,width=88mm}
\caption{
Stellar evolutionary tracks near the minimum mass $\mmin$
which  undergoes
a blue loop in the instability strip. The metallicity is indicated
 in the panels and
increases from the upper to the lower panel. Masses are indicated
along the tracks. Solid
lines correspond to the minimum mass, whereas
dashed lines correspond to stars which show a reduced blue loop
but do not cross the instability
strip. The crosses indicate
fundamental unstable modes, including the first crossing unstable
modes.}
\label{fig1}
\end{figure}

As shown in \S 3,  determination of PL relationships
is affected by the behavior of $\mmin$, since the reduction of the blue loop
affects the width of the instability strip at short periods. Indeed,
the blue edge (BE) is in this case determined by the turnover of the 
evolutionary track. Note that determination of blue edges
based on stability analysis of envelope models constructed
for a given mass - luminosity relationship and obtained
by varying arbitrarily $\te$ would predict
a hotter BE for masses near $\mmin$.
To illustrate this effect, we note that the fundamental  BE determined by the
blue loop extension of the 3 $\msol$ star with $Z$=0.004 is at $\te =
5860$ K, corresponding to $P_0$ = 1.2 days and $\log L/L_\odot$ = 2.384.
However, for the same mass, luminosity and chemical composition,
stability analysis performed on envelope models show that
fundamental modes should be unstable up to $\te$ = 6600 K and
$P_0$ = 0.83 days.

\begin{figure}
\psfig{file=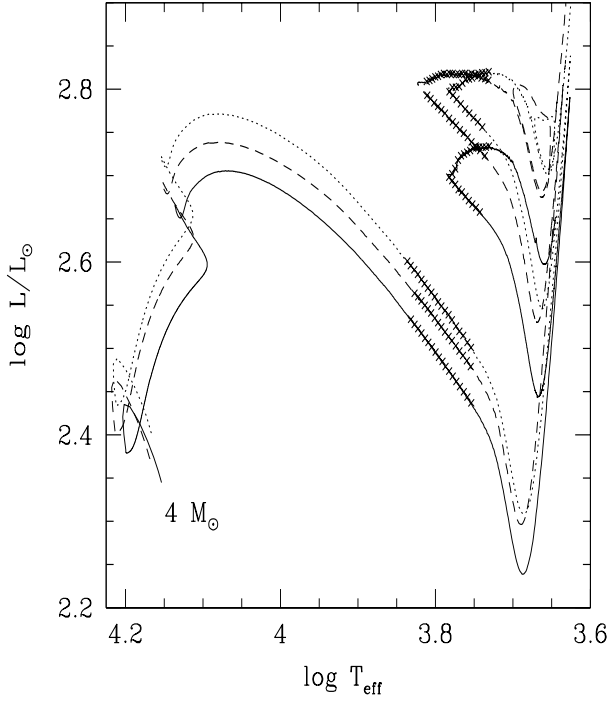,height=110mm,width=88mm}
\caption{
Evolutionary tracks of 4 $\msol$ with
different chemical compositions representative of the LMC: 
$(Z,Y)=(0.01,0.25)$ (solid curve), $(Z,Y)=(0.008,0.25)$
(dashed curve) and $(Z,Y)=(0.01,0.28)$ (dotted curve).
 The crosses indicate
fundamental unstable models.
}
\label{fig2}
\end{figure}

The extension of the blue loop  depends not only on the
metallicity but also on the helium abundance. This is shown in
Fig. 2
which displays the evolutionary tracks of 4 $\msol$ stars with
respectively $(Z,Y)=(0.01,0.25)$ (solid curve), $(Z,Y)=(0.008,0.25)$
(dashed curve) and $(Z,Y)=(0.01,0.28)$ (dotted curve). 
Since the size of the blue loop increases when $Z$ decreases
for a given mass near $\mmin$, we find $\mmin$ = 3.75 $\msol$
for $(Z,Y)=(0.008,0.25)$, instead of 3.875 $\msol$ for $Z$=0.01
(independently of $Y$).

The final important effect of chemical composition to be
mentioned in the present context concerns  the mass - luminosity
relationships.
Since for decreasing metallicity, the luminosity during core
helium burning increases for a given mass, this effect has
to be taken into account in order to correctly analyse the
metallicity dependence of PL relationships. 
 Figure 3 shows the ML relationships based on
the present standard calculations during the first crossing phase
and during the blue loop from $\mmin$ to 10 $\msol$\footnote {Above
10 $\msol$, the $Z$=0.004 and $Z$=0.01 models do not undergo
a blue loop. The 12 $\msol$ star with $Z$=0.02 does not undergo a blue
loop and reaches a lower $L$ during core He burning than the 11
$\msol$
star with the same $Z$.}. 
To estimate the following relationships,
we have taken simple means of L in the instability strips
(of F, 1H and 2H modes).
 We stress above all that these relationships  
illustrate the metallicity effect; more accurate
values are
given in the tables at the end of the present paper (Tables 6-8). 
 During the first crossing phase we find the following relationships:

\beq
\log {L \over L_{\odot}} = 3.21 \log { M \over M_\odot } + 0.51 \qquad\mbox{for}\qquad Z=0.02
\eeq
\beq
\log {L \over L_{\odot}} = 3.23 \log { M \over M_\odot } + 0.54 \qquad\mbox{for}\qquad Z=0.01
\eeq
\beq
\log {L \over L_{\odot}} = 3.28 \log { M \over M_\odot } + 0.63 \qquad\mbox{for}\qquad Z=0.004
\eeq

and in the blue loop IS:

\beq
\log {L \over L_{\odot}} = 3.55 \log { M \over M_\odot } + 0.53 \qquad\mbox{for}\qquad Z=0.02
\eeq
\beq
\log {L \over L_{\odot}} = 3.44 \log { M \over M_\odot } + 0.66 \qquad\mbox{for}\qquad Z=0.01
\eeq
\beq
\log {L \over L_{\odot}} = 3.43 \log { M \over M_\odot } + 0.85 \qquad\mbox{for}\qquad Z=0.004
\eeq

Note that the mean luminosity of SMC composition models
is   brighter than  galactic composition
models by $\Delta \log L/L_{\odot} \sim$ 0.2.
The initial
helium abundance affects as well the ML relationship, and an
increase of $Y$ from 0.25 to 0.28 for the $Z$=0.01 sequences yields
an increase in log $L/L_{\odot}$ by $\sim 0.1$. 

\begin{figure}
\psfig{file=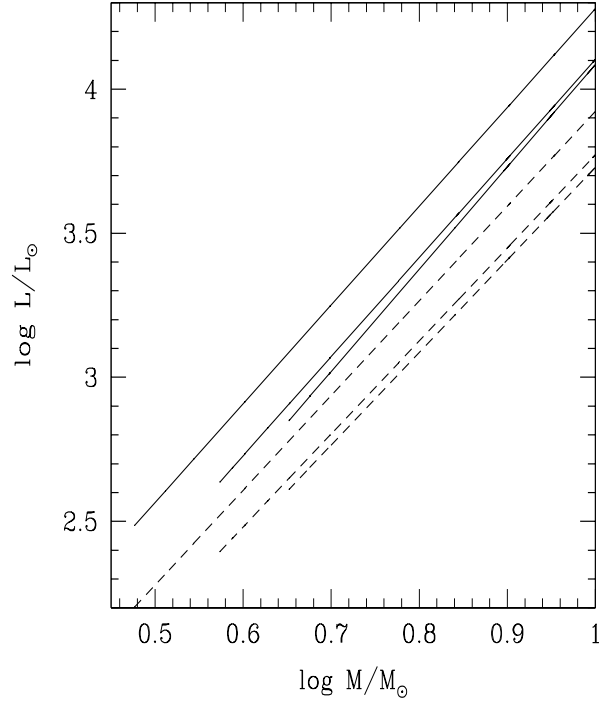,height=110mm,width=88mm}
\caption{
Mass - Luminosity relationships as a function of chemical composition.
The dashed curves correspond to the mean luminosity during the
first crossing instability phase and the solid curves
correspond to the blue-loop phase. The metallicity decreases
from top to bottom  and corresponds respectively to:
$(Z,Y)=(0.004,0.25)$ (upper curve), 
$(Z,Y)=(0.01,0.25)$
(middle curve) and 
$(Z,Y)=(0.02,0.28)$ (lower curve).
}
\label{fig3}
\end{figure}

\subsection {Evolutionary timescales}

The determination of PL relationships and the probability
to detect a Cepheid of a given mass in a large sample of data will
heavily depend on the time spent in the instability strip. Regarding
the first crossing, this time is fixed by the thermal
timescale of the star $\tau_{th} = Gm^2/RL \sim  0.8 (m/\msol)^2 \te^2
(L_\odot/L)^{3/2}$ yrs. $\tau_{th}$ varies typically from $\sim 10^{5}$ yrs
for 3 $\msol$ to $\sim 10^{4} - 5 \, 10^{3}$ yrs  for 8-10
$\msol$. The time spent in the instability strip during the first
crossing is even shorter than $\tau_{th}$ and is roughly
100 times shorter than the time spent in the blue loop instability
strip.    

Regarding this latter timescale, Fig. 4 displays evolutionary tracks
of stars of different masses and indications of the time spent
during the 
core He burning phase. 
The distribution of the evolutionary timescales in this figure is
generic of
core helium burning during a blue loop and remains valid 
independently of metallicity or convection treatment.
As shown in Fig. 4, most of the time during core He burning phase is spent 
near the turnover of the evolutionary tracks. The first branch of
the blue loop
is characterized by an increase in mass of the He convective core,
 the expansion
of the regions inner to the H burning shell and
the contraction of the envelope: evolution proceeds toward higher
$\te$. As the central nuclear
fuel (He) is depleted, the central regions start to contract
again, resulting in an expansion of the envelope, and the evolution
proceeds toward lower $\te$ on a shorter timescale 
determined by the faster depletion of He.
Interestingly enough, the mean position of a given mass in the IS, taking
into account its evolutionary time, is roughly located in
the middle of the IS. Thus the time distribution in the IS
does not favor a particular position near the IS 
edges, except for masses near $\mmin$ for obvious reasons
(cf. \S 3.2 and Fig. 7). 
This result is valid for all metallicities studied and 
independently of the convective treatment.


\begin{figure}
\psfig{file=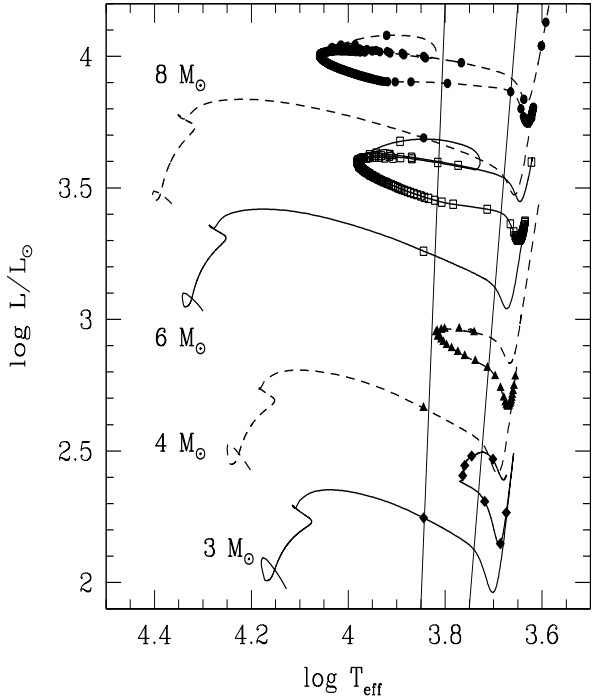,height=110mm,width=88mm}
\caption{
Evolutionary tracks for stars of different masses, as indicated in the
figure, for $Z$=0.004. Symbols are plotted along each tracks every
$\Delta t$, with $\Delta t = 10^7$ yrs (full diamonds, 3 $\msol$),
$\Delta t = 10^6$ yrs (full triangles, 4 $\msol$),
$\Delta t = 10^5$ yrs (open squares, 6 $\msol$) and
$\Delta t = 2 \, 10^4$ yrs (full circles, 8 $\msol$).
The time count starts on the first crossing at $\te$ = 7000K.
The vertical lines indicate roughly the location of the instability
strip. 
}
\label{fig4}
\end{figure}

\subsection{Uncertainties due to the treatment of convection}

In order to estimate the uncertainties of PL relationships relying
on stellar evolution calculations, we have analysed the sensitivity
of the results to a variation of the mixing length parameter $l_{\rm mix}$
and to the effect of overshooting or any enhanced mixing in the core.
The effect of overshooting on stellar models and  the necessity
to include it or not in order to fit
observations has been
 already widely investigated (see Bressan et al. 1993;
El Eid 1994; Pols et al. 1998; and references
therein) and is out the scope of the present paper. 

An increase of $\alpha_{\rm mix}$ from our standard value 1.5 to 2 results
essentially in the well-known shift of the Red Giant Branch to higher
$\te$. However, the ML relationships and the aspect of
the blue loops remain essentially unchanged. We will  see in
\S 3 that this parameter affects the stability of
the models and modifies the width of the IS.

The effect of enhanced mixing in the core as modeled by our simple
recipe is more important regarding stellar evolution, and
results in an increase of the luminosity for a given mass by
0.15 - 0.20 in $\log \, L/\lsol$ for our adopted overshoot
length $d_{\rm ov} = 0.15 H_{\rm P}$. This is a little less than the mild-overshooting
used by Chiosi et al. (1993) with an increase of the luminosity
$\log (L/\lsol)_{\rm ov} = \log (L/\lsol)_{\rm standard} + 0.25$. The size of
the blue loops are affected by overshooting and most
importantly
the minimum mass $\mmin$ which undergoes a blue loop 
increases, in agreement with similar calculations including
overshooting (see El Eid 1994 for
a detailed discussion). 

As discussed in Baraffe et al. (1998), the standard evolutionary
models with $Z$=0.01 do not predict the right position
of the faintest Cepheids
observed in the EROS-1 and EROS-2 LMC samples,
which show an abrupt end of the PL relationship for fundamental
mode pulsators at a
period $P$ = 2.5 days (Sasselov et al. 1997; Bauer et al. 1998).
This observational trend has been confirmed very recently by the MACHO survey
(Alcock et al. 1998).
Although few objects are observed below this period,
the bulk of fundamental mode Cepheids clearly do not
extend below this value. 
One may interpret this abrupt end in terms of the faintest unstable models
on a blue loop, since most of the time is spent during this phase.
The few objects below 2.5 days may then correspond to first crossing
models and their scarcity is representative of the faster evolutionary
timescale of this phase (cf. \S 4).

Our standard calculations for $Z$=0.01 predict this abrupt end
at $P \sim$ 1.8 days, corresponding to the blue loop unstable models
of $\mmin$ = 3.875 $\msol$. A possible explanation for this discrepancy
is a shortcoming in the stellar evolutionary models which may
underestimate $\mmin$.
With our overshooting prescription and an overshoot distance fixed at
0.15 $H_{\rm P}$ the minimum mass $\mmin$ is shifted to 
4.25 $\msol$ instead of 3.875 $\msol$ for $Z$=0.01. As shown in \S 4,
this yields  better agreement with the lower end of the PL relationship
for LMC fundamental mode Cepheids.
When applying the same overshooting prescription to $Z$=0.004
models ({\it i.e.}, the SMC metallicity), $\mmin$   
is shifted to 3.25 $\msol$,
instead of 3 $\msol$ in the standard case. However, for the SMC, Baraffe et al.
(1998) show that the standard models are in excellent agreement with
the observed distribution of fundamental mode Cepheids, and overshooting
is  not required in this case (cf. \S 4). 
But the idea of invoking some kind of extra-mixing for LMC evolutionary
models but not for the SMC composition in order to improve the
comparison with observed fundamental Cepheids is not satisfactory. Our poor
understanding
of core overshooting in the current stellar evolution theory does not
allow any predictions or speculations regarding the efficiency of such
mixing as a function of metallicity. 
We will come back to this problem in \S 4.
 
\section{Theoretical Period-Luminosity relationships}

\subsection{Linear stability analysis}

In order to get fully consistent calculations between evolutionary
and pulsation calculations, a linear non-adiabatic stability analysis is
performed directly on the complete evolutionary models along the
tracks.
 This provides fully
consistent mass-luminosity-period-evolutionary time relationships
and enables us to analyse carefully evolutionary effects on the PL relationships.
The pulsation calculations are performed with a radial pulsation 
code originally developed by Umin Lee (Lee 1985).
 The main uncertainty inherent to the
stability
analysis concerns the pulsation-convection coupling, which is presently
neglected, {\it i.e.}, the perturbation $\delta F_{\rm conv}$ of the
convective
flux is neglected in the linearized energy equation. Such a crude
approximation, adopted also in Saio and Gautschy (1998),
is resorted to our poor knowledge of the interaction
between convection and pulsation, which requires
a time-dependent non-local theory of convection. This
interaction provides a damping term which is necessary to
 obtain a red edge for the Cepheid
IS. Perturbed models which assume $\delta
F_{\rm conv}$ = 0 (cf. Fig. 5) remain unstable even near the Red Giant Branch
(cf. Saio and Gautschy 1998, Bono et al. 1998a; Yecko et al. 1998).
 
Several attempts have been made to take into account
pulsation-convection
coupling in Cepheids with the most recent 
approach
 by Yecko et al. (1998; see also Kollath et al. 1998)
and BCCM98. A systematic survey of linear properties
of Cepheid models including a model of turbulent convection
performed by Yecko et al. (1998) shows the extreme sensitivity
of the instability strip location on the choice of the several free parameters
inherent to the model of convection. As demonstrated by
Yecko et al. (1998), the location and width of the IS depend
essentially on the three parameters which 
determine the mixing length,
the convective flux and  the eddy viscosity. 
The only
possible method to calibrate these parameters is a comparison with
observed IS. 
However, one can expect  that a theoretical determination
of the IS depends not only on pulsation calculations, but also on
stellar evolution properties. 

Since the purpose of the present paper is to determine
to which extend evolutionary properties are important for the position
and width of the IS, we have chosen in a first study
to limit the range of free parameters in the  pulsation calculations,
and to adopt a rather arbitrary criterion for the determination
of the IS red edge.
A forthcoming paper will be devoted to the uncertainties resulting
from the treatment of convection in the pulsation calculations. 

In the present models, as shown in Fig.5, 
we assume ad hoc a red line corresponding to
 the effective temperature
where the 
  growth rate reaches a maximum  as $\te$ decreases
along the evolutionary
track for a given mass. This method is the same
as adopted in Chiosi et
al. (1993).
 Adopting a time dependence
for the eigenfunctions of the form $\exp (i\sigma_k t)$ where
$\sigma_k = \sigma_r + i \sigma_i$ is the eigenfrequency
of the $k_{th}$ mode,
 the growth rate is defined as $\eta_k = -
\sigma_i/\sigma_r$. Thus, positive
 growth rates indicate unstable modes.

\begin{figure}
\psfig{file=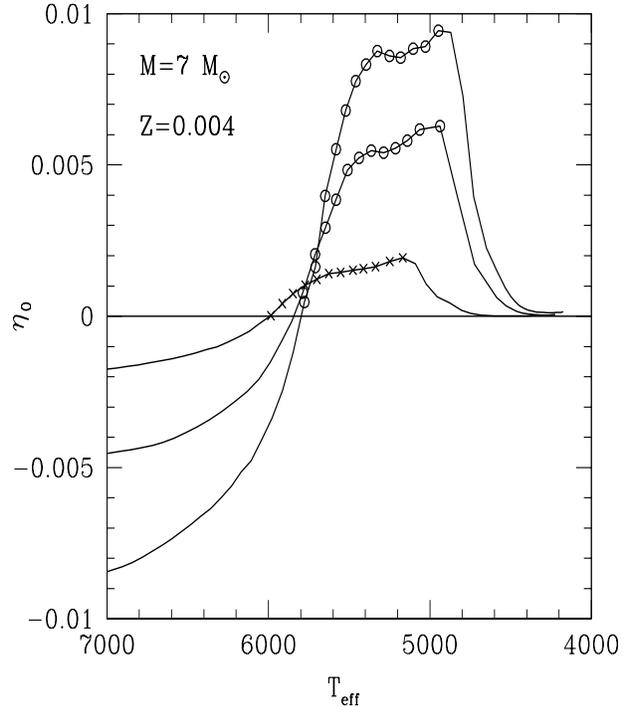,height=110mm,width=88mm}
\caption{
Evolution of the fundamental mode growth rate $\eta_0$ as
a function of effective temperature along the evolutionary track
of a 7 $\msol$ with $Z$=0.004. The selected
unstable
modes which define the IS are indicated by crosses on the first
crossing  and by full circles during the
blue loop phase. According to the criterion adopted in the present paper, 
the red edge is given by the location of the maximum
reached by $\eta_0$ as $\te$ decreases.
}
\label{fig5}
\end{figure}

\subsection{Effect of chemical composition}

Since most of the theoretical analysis devoted to Cepheids 
are based on stability analysis of envelope models  
and ML relationships independent of  metallicity, it
is interesting to analyse the uncertainties resulting from such
assumptions. We have tested models  by comparing results obtained with the
full
computations (stellar evolution and pulsation calculations coupled) and 
with envelope models constructed with 
constant luminosities. Fig. 6a  shows the evolutionary tracks of
 7 $\msol$ and 11 $\msol$  stars with
$Z$=0.004 and $Z$=0.02 and the fundamental unstable  models during
core He burning phase. The luminosities adopted for the envelope
calculations are derived from our fits given in \S 2.1
and are indicated by full circles for $Z$=0.02 (Eq. (4)) and full
squares for $Z$=0.004 (Eq. (6)). The results of the linear stability
analysis are shown in Fig. 6b . The differences between the IS location
based on the full calculations and envelope models are small
when using the metallicity dependent ML relationships.
The main differences are due to  differences
in the luminosity or to departure from the initial composition
in the case of full calculations, as a consequence of the first
dredge-up. As an example, we note that the red edge given by stability
analysis
of envelope
models  for 11 $\msol$ with $Z$=0.02 (full circles) is slightly
redder by $\sim$ 100 K than the red edge given by the full
calculations (solid curves). This is mostly due to an enrichment of He in
the envelope of this star from the initial value $Y$=0.28 to $Y$=0.31.
When adopting the latter value to construct the envelope models, 
the red edge is shifted to higher $\te$, in better agreement
with realistic models. However, we note that such evolutionary
effects affecting the initial composition are small
and  yield to a shift of the
red edge of at most 100 K in $\te$. We also stress  that this
result highly depends on our rather uncertain criterion for
the red edge determination.  

The most important effect is obtained when adopting
for the $Z$=0.02 models the same ML relationship as derived
for $Z$=0.004 (Eq. (6)). The main result is
a shift of the IS toward lower $\te$ (full triangles)
and to a larger width, with a shift of the red edge by 100 to 200 K,
compared to the $Z$=0.004 case (dashed lines and full squares).
This example highlights  the metallicity dependence of the ML relationship.

\begin{figure}
\psfig{file=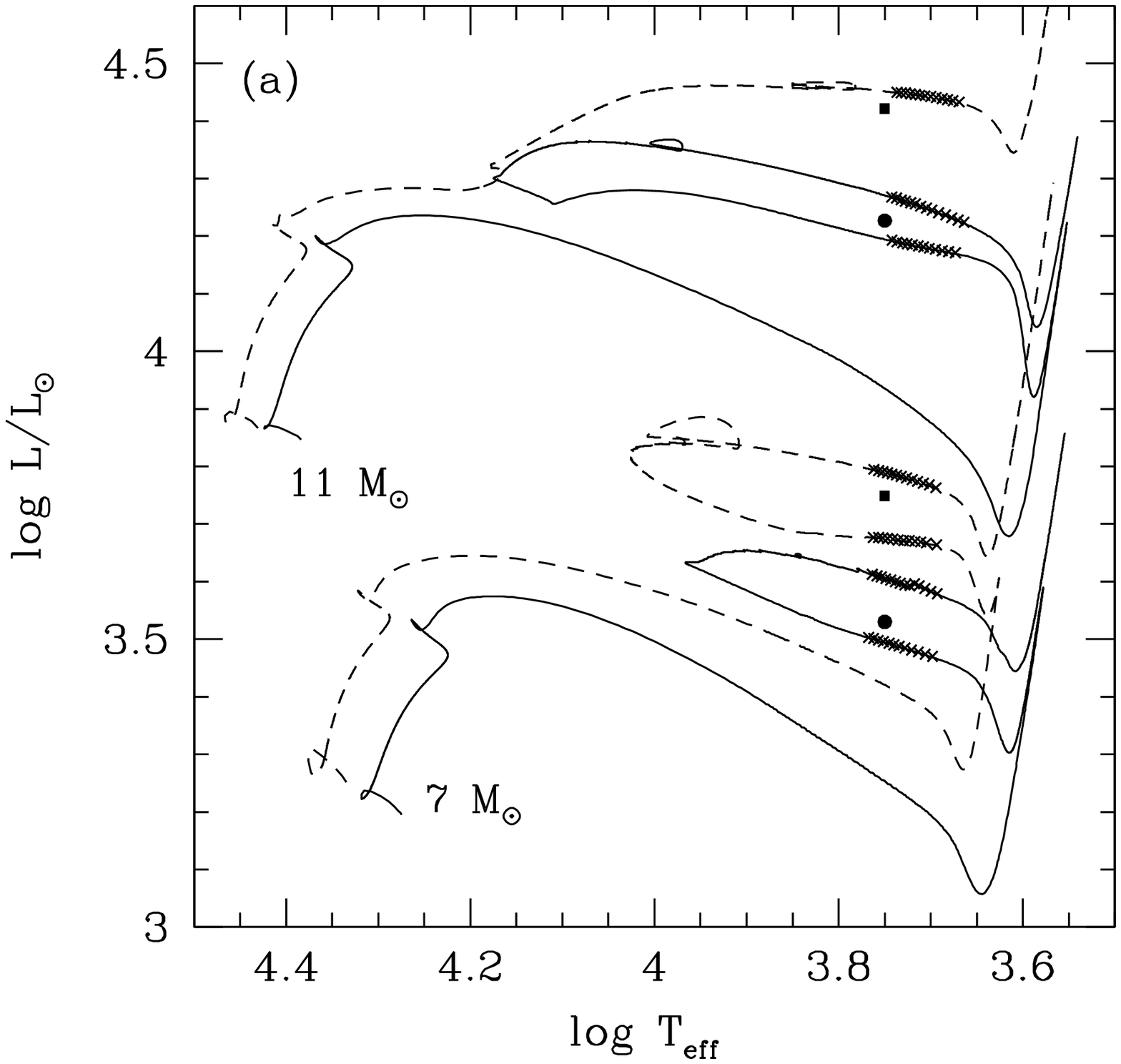,height=88mm,width=88mm}
\psfig{file=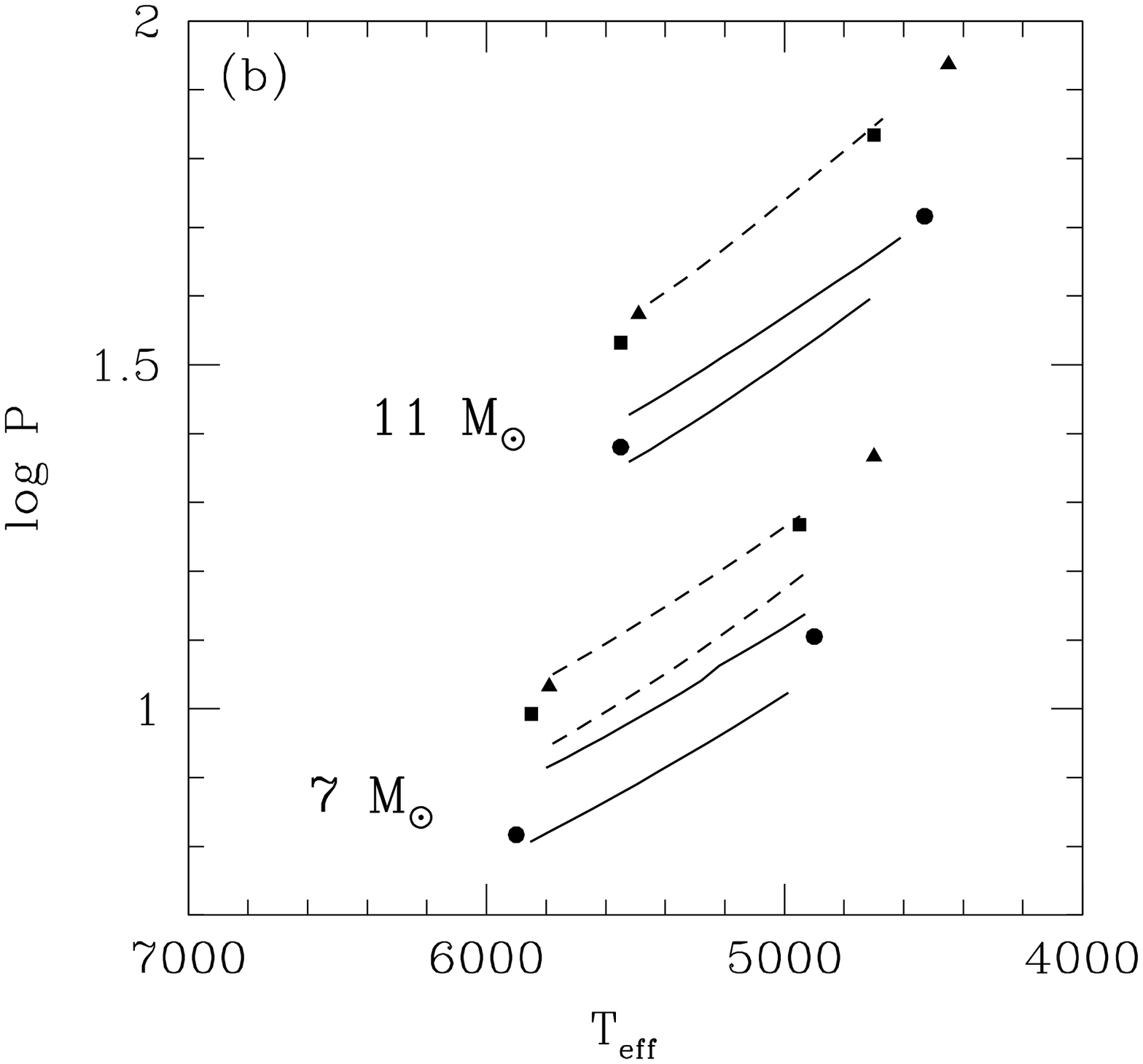,height=88mm,width=88mm}
\caption{
{\bf (a)} Evolutionary tracks for stars of 7 $\msol$ 
and 11 $\msol$, as indicated in the
figure, for $Z$=0.02 (solid curves) and $Z$=0.004 (dashed curves). The
location of fundamental unstable
modes during core He burning is indicated by crosses.
The full circles indicate the luminosity given by the ML
relationship of Eq. (4) for 7 $\msol$ and 11 $\msol$ and the full
squares correspond to Eq. (6). {\bf (b)} Period (in days) 
of the fundamental mode as
a function of $\te$ for the same models as in (a).
 The curves correspond to  coupled evolutionary
and pulsation calculations with same masses and
metallicity as in (a). The symbols correspond to the
blue and red edges obtained from stability analysis of  envelope models 
with composition $(Z,Y)=(0.02, 0.28)$ and $L$ given by Eq. (4) (full
circles)
and $(Z, Y)$=(0.004, 0.25) and $L$ given by Eq. (6) (full squares).
 The full triangles correspond to $(Z,Y)$=(0.02, 0.28)
envelope models with $L$ given by  Eq. (6), i.e., the same luminosity
as for the $Z$=0.004 models.
}
\label{fig6}
\end{figure}

The instability strips  in period-luminosity and period - $\te$ 
diagrams based on
our standard models in the core He burning phase 
are shown Fig. 7 for $Z$=0.004, 0.01 and
0.02. In agreement with the results of Saio and Gautschy (1998),
we do not find significant metallicity effects on the location
and width of the IS. We only note that for $\log \, P > 1$, the
red edge of the solar metallicity IS moves toward slightly cooler
$\te$. Because of the arbitrariness of our red edge determination
and the increasing influence of convection in such models, this
result is to be taken with caution. The lower panel of Fig. 7
indicates the mean location in the IS of each mass,
taking into account its time spent at
a given location in the IS \footnote {The mean position is determined
from a weighted mean {\it i.e} for a given model {\it i} with age
{\it t(i)} in the IS, the quantities 
$L(i)$, $\te(i)$, etc..., are weighted by the time interval
$t(i+1) - t(i)$.}.   
Except for the particular case
of $\mmin$, indicated by arrows in Fig. 7, the mean position
given by such weighting function
for each stellar mass is roughly located in the middle of the IS,
and shows the negligible dependence on $Z$. 
Such results do not support the conclusions of Fernie (1990) who
find a non uniform population of Cepheids in the IS. The discrepancy
between theoretical predictions and such observations has
been recently discussed by  Gautschy (1999), and is confirmed by the present
work. A more detailed analysis is under progress to 
determine whether the Fernie (1990) results can be explained by an observational bias or by intrinsic properties of stellar models which escape to
the present analysis.  

\begin{figure}
\psfig{file=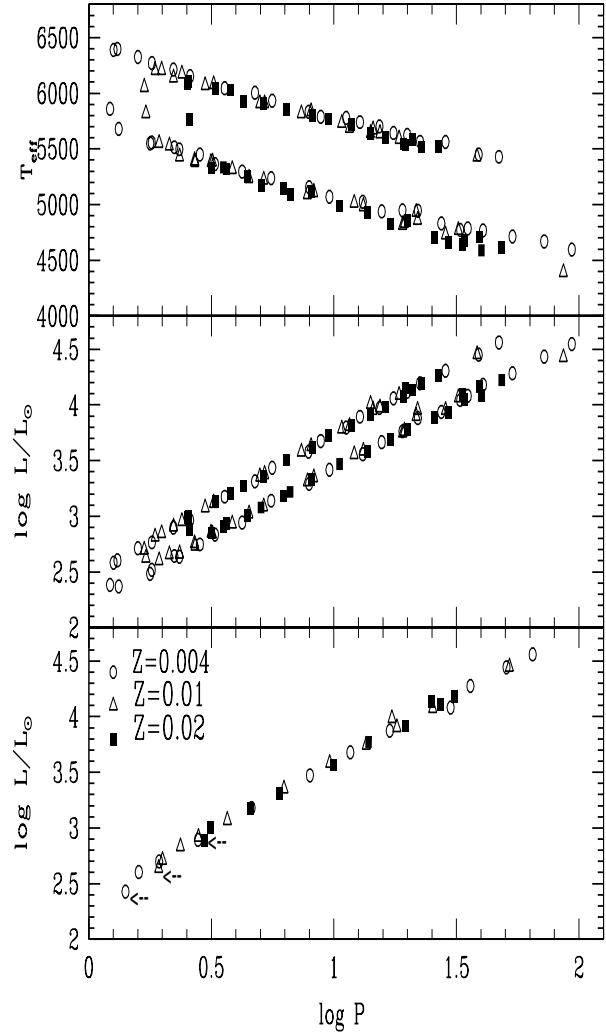,height=160mm,width=88mm}
\caption{$P$ - $L$ (middle panel) and $P$ - $\te$ (upper panel)
 diagrams for the blue and red edges of the 
fundamental modes during the core He burning phase of 
standard models for three metallicities.
 The lower panel indicates the mean location
in the IS of each mass, taking into account the evolutionary times
(see text). The arrows indicate the mean position
of $\mmin$ for each metallicity displayed.
 The symbols correspond to different metallicities
(cf. lower panel). 
}
\label{fig7}
\end{figure}

Metallicity however affects the lower end of the period distribution
since it modifies $\mmin$ (cf. Fig. 1). As suggested in Baraffe et al.
(1998), the reduction of the He blue loop yields a change of slope
in the PL relationship. Because of the metallicity dependence of
$\mmin$, the period at which this change of slope
is predicted decreases with metallicity. As illustrated in Fig. 8
for standard models, this change of slope takes place at log $P \sim$
0.2 for $Z$=0.004, log $P \sim$ 0.35 for $Z$=0.01
and log $P \sim$ 0.5 for $Z$=0.02. Although this value may
change due to uncertainties in the convection treatment of
 the {\it evolutionary
models}, as to be mentioned in the next section, the relative effect due to
the metallicity is unaffected by such uncertainties.
Fig. 8 displays the unstable modes during the first crossing,
which are expected to be observed in a sufficiently large
sample of data. As already mentioned, the time spent during this phase
is $\sim$ 1/100 the time spent in the blue loop. Thus in a large
sample
of data, one expects to see the trend of a change of slope followed
by most of the low period Cepheids, and a few of them with even lower periods 
which do not show a change of slope and belong
to the first crossing. Note that, although the time spent in the first
crossing is relatively short, it remains
too long to observe any variation of the period on a timescale of
years. Indeed, for a 4 $\msol$ with LMC composition, we predict a variation
of $P_0$ during the first crossing of less than $10^{-5}$ days
per year. 

Finally, we derive PL relationships based on the following method: 
we assign   a mean position to {\it each} stellar mass, according
to the time  spent in different locations in the IS,
 and derive a {\it linear}
least-square fit to these points from $\mmin$
to 12 $\msol$. In the following, all relationships are based
on this method, 
which is more convenient for an analysis
of metallicity effects and 
a comparison with different sample of data over
a wide range of periods, than the one adopted
by Baraffe et al. (1998), which gives higher weight
to the lowest mass stars and favor the low period region.  
%
%
%

Table 1 summarizes the coefficients of the present PL
relationships as a function of metallicity
and convection treatment.
An inspection of Table 1 shows a negligible dependence of
the PL relationships on Z. 
As expected from Fig. 8, the PL relationships have
much steeper slopes  near $\mmin$:  we obtain a slope of 1.36 from
$\mmin$ = 4.75 $\msol$ to 5.5 $\msol$ ($\log P \, \simle 0.8$)
for $Z$=0.02, 1.46
from $\mmin$ = 3.875 $\msol$ to 5 $\msol$ ($\log P \, \simle 0.7$)
for $Z$=0.01 and
1.47 from $\mmin$ = 3 $\msol$ to 4 $\msol$ ($\log P \, \simle 0.6$)
for $Z$=0.004. However, PL relationships derived 
from respectively 5.5, 5 and 4 $\msol$ for $Z$=0.02, 0.01 and 0.004,
 up to 12 $\msol$, yield essentially the same results as obtained
in Table 1. 
This indicates that a unique linear PL relationship
is not appropriate on the whole range
of periods involved in the IS, and that PL relationship
based on {\it low period} Cepheid samples must be taken with caution.
Finally, we note that a slight variation of the helium
abundance from 0.25 to 0.28 with $Z$=0.01, or a slight
variation of $Z$ from 0.01 to 0.008 with $Y$=0.25, yields
negligible effects on the PL relationships. 

Comparison between our blue edges and the one obtained by Saio and Gautschy
(1998) shows a {\it constant} shift of their IS toward brighter
L by $\Delta \log \, L \sim 0.1$
for a given $P$.
This explains the different PL relationships
 between their work and the present one.
Their red edge is determined by shifting
the blue edge by $\Delta \log \, \te = 0.06$, which differs from
the method adopted in the present work, but yields a width of
$\sim$ 700 K in agreement with the present work.
The reason of the difference of $\Delta \log \, L \sim 0.1$ is
not clear, but is certainly acceptable, since
different codes are used.
Saio and Gautschy (1998) did not analyze  the low mass region
where the blue loop extension is reduced, since their analysis
is performed for m $\ge 4 \msol$ for all metallicities. 
It is worth  mentioning
that our two independent studies reach  
 the same conclusions  regarding the small effects of metallicity
on PL relationships. 

\begin{figure}
\psfig{file=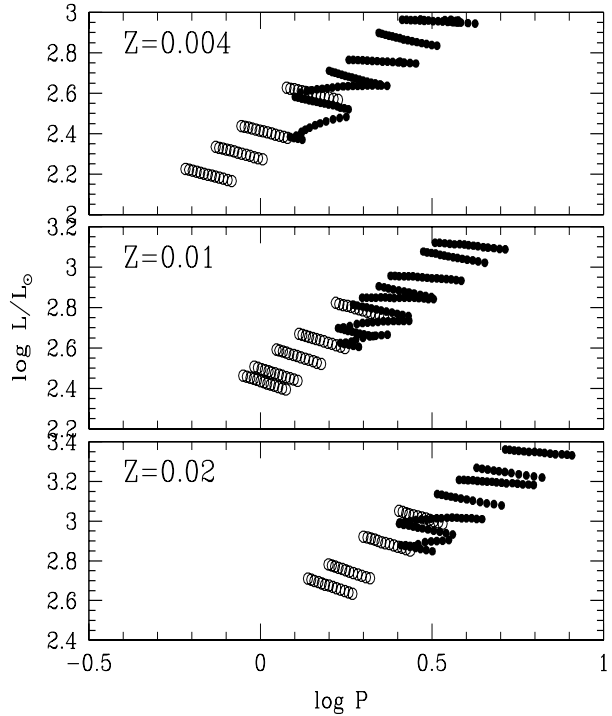,height=110mm,width=88mm}
\caption{PL diagram for fundamental modes during the first
crossing (open circles) and the blue loop
phase (full circles) of standard models as a function of metallicity.
The corresponding masses shown in each panel are : 3, 3.25, 3.5 and 4 $\msol$
for $Z$=0.004 (upper panel); 3.875, 4, 4.25, 4.5 and 5 $\msol$
for $Z$=0.01 (middle panel); 4.75, 5, 5.5 and 6 $\msol$ for $Z$=0.02 (lower
panel)
}
\label{fig8}
\end{figure}

\subsection {Effect of different treatments of convection}

In this section we analyse the effect of a variation
of $l_{\rm mix}$ and of overshooting on the location
of the IS and on the PL relationships. 
An increase of $l_{\rm mix}$ from 1.5 to 2 in the evolutionary 
calculations affects  essentially  the location of
the red edge, since the structures of these  models are mostly affected
by convective transport in their envelope. An increase of $l_{\rm mix}$
yields lower temperature gradients
in the region of H and He partial ionization and results
 in a denser structure for a given
$T$. The net result is a damping effect and a shift of the
red edge by $\Delta \, \log \te \sim 0.02$ toward hotter $\te$.
However, an inspection of Table 1 shows that the effect
on the PL relationships is small.

\begin{table}
\caption{Coefficients of  $\log \, P - \log \, L/L_\odot$
 relationships (slope, zero-point)
for fundamental pulsators as
a function of metallicity and convective treatment in the evolutionary
models. The standard models refer to $(\alpha_{\rm mix}, d_{\rm ov}) = (1.5, 0)$,
{\it i.e.} $\alpha_{\rm mix}$ = 1.5 and no overshooting.}
\begin{tabular}{lccc}
\hline\noalign{\smallskip}
($\alpha_{\rm mix}, d_{\rm ov}$) & (1.5,0) &(2,0) &
(1.5,0.15)    \\
\noalign{\smallskip}
\hline\noalign{\smallskip}
$Z$=0.02  & (1.237, 2.347) &    -   & - \\
$Z$=0.01  & (1.248, 2.346)  & (1.256, 2.386) &  (1.270, 2.264) \\
$Z$=0.004 & (1.239, 2.328)   & (1.271, 2.358)  & (1.289, 2.265) \\
\hline
\end{tabular}
\label{edge1cz43}
\end{table}

Overshooting  does not significantly modify the location
of the IS. As shown in Table 1, overshooting yields slightly different 
PL relationships compared to the standard case, yielding
a maximum of  15\% difference in $L$ at a given $P$. 
However the effect remains  small on the whole range of periods investigated, 
in agreement with Saio and Gautschy (1998). 
The previous conclusions regarding the metallicity effects
remain valid when overshooting is taken into account.

As mentioned previously, models including overshooting
predict a change of slope at the lower end
of the PL relationship at {\it higher} $P$, compared
to the standard case. As shown in Fig.9,
this change of slope is expected at log $P \sim$ 0.5 for $Z$=0.004 and
log $P \sim$ = 0.65 for $Z$=0.01.
As for the standard models, derivation of PL relationships
on a  mass range near $\mmin$ yields much steeper slopes as well.
 
\begin{figure}
\psfig{file=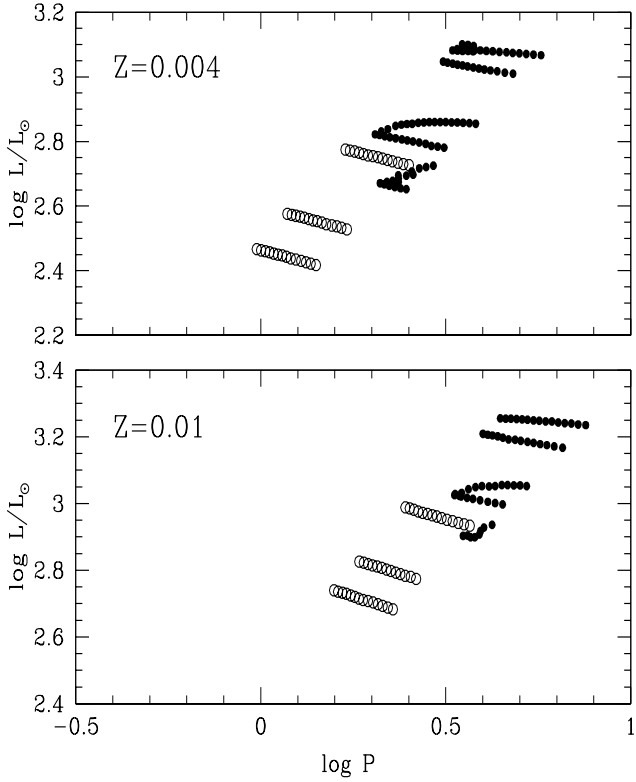,height=110mm,width=88mm}
\caption{Same as Fig. 8 for models including overshooting
with $d_{\rm ov} = 0.15$.
The corresponding masses shown in each panel are: 3.25, 3.5 and 4 $\msol$
for $Z$=0.004 (upper panel);  4.25, 4.5 and 5 $\msol$
for $Z$=0.01 (lower panel).
}
\label{fig9}
\end{figure}

\subsection{First overtone pulsators}

The main effects of chemical composition and convection
treatment on the PL relationships remain the same for
first overtone pulsators. We thus do not find significant effect
of $Z$ on the location of the 1H IS and the PL relationships.
Table 2 summarizes the PL relationship
coefficients as a function of metallicity for the standard
models. 

\begin{table}
\caption{Same as Tab. 1
for first overtone pulsators and standard models.}
\begin{tabular}{lc}
\hline\noalign{\smallskip}
($\alpha_{\rm mix}, d_{\rm ov}$) & (1.5,0) \\
\noalign{\smallskip}
\hline\noalign{\smallskip}
$Z$=0.02  & (1.245, 2.661)  \\
$Z$=0.01  & (1.281, 2.618)  \\
$Z$=0.004 & (1.258, 2.611) \\
\hline
\end{tabular}
\label{edge1cz43}
\end{table}

The minimum mass $\mmin$ undergoing
a blue loop in the {\it fundamental mode} IS does not fulfill the instability
criterion for 1H, because of its hotter $\te$
compared to F IS. Thus, $\mmin$ increases for 1H and
corresponds to 3.25 $\msol$ for $Z$=0.004, 4 $\msol$ for
$Z$=0.01 and 5 $\msol$ for $Z$=0.02. We predict as well  
the feature of a change of slope at the lower end
of the PL relationship, as shown in Fig. 10. However the effect seems
 to be less pronounced, because of the significant
reduction of the IS width, by more than 400K, compared to F modes.
 We are aware that this
result strongly depends on our adopted criterion  to
determine the red edge, which fixes the width of
the IS.
This point is further analysed on the basis of comparisons 
with observations in \S 4.
  
\begin{figure}
\psfig{file=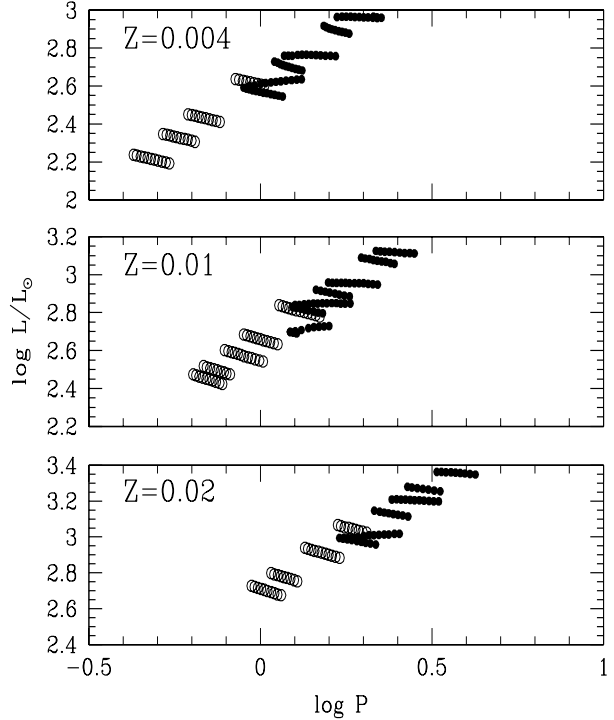,height=110mm,width=88mm}
\caption{Same as Fig. 8 for first overtone pulsators. For the first crossing
phase, the same masses as in Fig. 8 are displayed. For the blue
loop phase, the minimum masses displayed are 
$\mmin$ = 3.25 $\msol$ for $Z$=0.004, 
$\mmin = 4 \msol$ for $Z$=0.01 and $\mmin = 5 \msol$ for $Z$=0.02.
}
\label{fig10}
\end{figure}

\section{ Comparison with observations}

\subsection{Synthetic spectra}

In order to perform comparisons with observed magnitudes and colors,
we have used version 10.3 of the model atmosphere code {\tt PHOENIX}
to calculate a set of static atmosphere models, and their synthetic
spectra, in local thermodynamic equilibrium (LTE), using a spherically
symmetric radiative transfer. The atomic and molecular opacities are
included in a direct line-by-line opacity sampling treatment (i.e. no
pre-tabulation) with a line selection criterion applied at each model
iteration and based on the strength of the line relative to the
strength of local continuum opacities.  Further details of the general
input physics are discussed in Hauschildt et al. (1998) and references
therein. The model grid englobes the parameter range of Cepheids, {\it
i.e.}, $\te = 4000 - 7000 K$ and $\log g = 0 - 3.5$, and is an
extension to lower gravity of the models recently developed by
Hauschildt et al. (1998), which successfully describe M-dwarf
atmospheres (cf. Allard et al. 1997).  Convective mixing is included
according to the Mixing Length Theory (MLT). Only models with a mixing
length of unity and a sphericity corresponding to a stellar mass of $M=5\,M_\odot$ were calculated.
We have checked that for the range of $\te$ and gravities involved in
the present study, the sensitivity of the synthetic spectra to the
stellar mass is small, justifying the above approximation. Only for
low gravity and effective temperature, the effect can be important and
reaches 10\% difference in models with $T_{\rm eff}=3600\,$K, $\log
\,g = 0.0$ and solar abundances. For higher gravities, this effect
vanishes, as expected.
The depth-independent microturbulence used in the model atmosphere
calculations is 2 km/s. Changes will affect high-resolution 
spectra far more than the broad band colors used here, so we have
not considered other values of the microturbulence.

In the following, 
the $BVRI$ magnitudes are based on the Johnson-Cousins system (Bessell
1990),
whereas the $JHK$ magnitudes are defined in the CIT system (Leggett
1992). 
Since some of the observations used in the present paper are based
on the SAAO-Carter system, we have also calculated magnitudes in
this system, using the filter transmission curves kindly provided
by P. Fouqu\'e. The $J$-band is mainly affected by the choice of
the photometrical system, whereas the $K$-fluxes in both systems
do not differ by more than 0.05 mag. 
 Comparison between our
results in CIT and Carter systems yields the following
conversion equation:

\beq
 (J-K)_{\rm CIT} = 0.91 \, (J-K)_{\rm Carter},
\eeq

\noindent adopting $K_{\rm CIT} = K_{\rm Carter}$.
This equation, which is accurate within less than 0.01 mag
for the range of $\te$ and
surface gravity of interest, is used in the following
 to transform in the CIT system
observations
given in the Carter system. We stress that Eq. (7) is {\it only
valid for giants} but fails for cooler and higher gravity
objects. 

\subsection{Period - Magnitude relationships}

This section is devoted to comparison with observations in period - magnitude
diagrams. 
The metallicity effects on the period - magnitude relationships
are analyzed and quantified in \S 5.

\subsubsection{Galactic Cepheids}

The solar metallicity models are compared to the sample of galactic
Cepheids of Gieren et al. (1998, GFG98). The careful selection of 
stars
in this sample and the determination with  good accuracy of their radius
and distance yield  lower dispersions of the derived period-magnitude
relationships
and thus provide an excellent test for theoretical models. 
Comparison
between models and observations are shown in Fig. 11 in the $V$, $J$ and $K$
bandpasses. The $J$-magnitudes of GFG98 are given in the Carter System,
and
have been transformed in the CIT system according to (7).

\begin{figure*}
\psfig{file=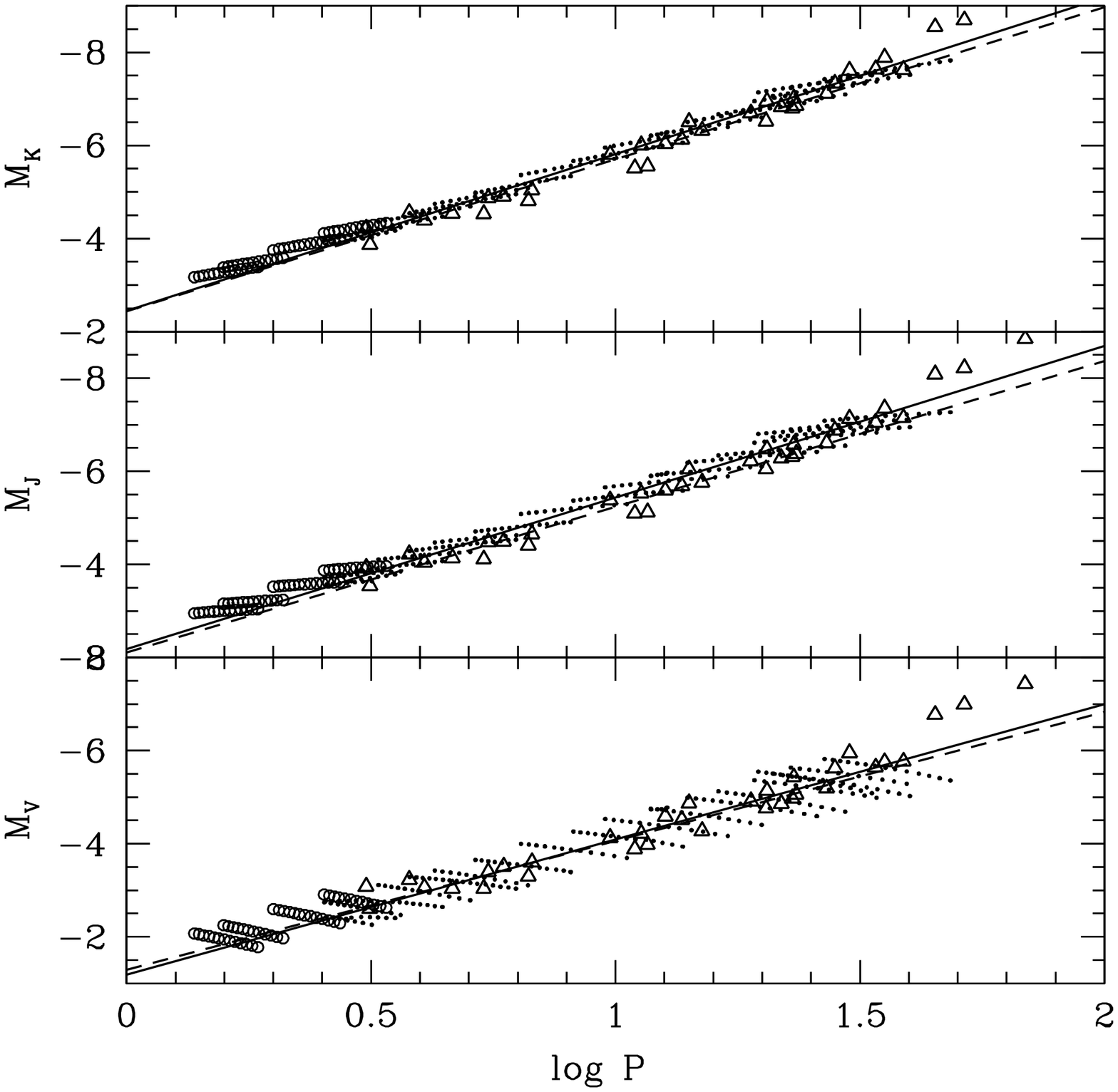,height=200mm,width=176mm}
\caption{Period - Magnitude diagrams in the $VJK$ bands for
solar metallicity standard models. The dots correspond to fundamental
unstable modes during core He burning phase from $\mmin$ = 4.75 $\msol$
to 12 $\msol$. 
The open circles correspond
to first crossing fundamental unstable modes for 
 4.75, 5, 5.5 and 6 $\msol$. Observations (open triangles) are from
GFG98. The dashed curves correspond to
the PL relationships given by GFG98. The solid curves correspond
to the relationships derived from the present standard models.
}
\label{fig11}
\end{figure*}

In a similar way as done for the PL relationships in \S 3, we derive
 period - magnitude relationships.
The coefficients of these
relationships are given in Tab. 3  for each $Z$ studied.
Fig. 11 shows a comparison between the predicted relationships and
those derived by GFG98 using  both galactic and LMC Cepheids.

Excellent agreement is found between models and data in the
three bandpasses, as well as between the theoretical and
observed mean PL relationships. 
%
The present
models provide a much better match to observations than the BCCM98
 models, based on non-linear calculations but not on consistent
evolution-pulsation calculations, and compared
to the same data by GFG98.
For galactic
Cepheids with $\log \, P > 1.3$, a better agreement in a $P - \mv$
diagram is reached with the $Z$=0.008 models of BCCM98,
whereas the $Z$=0.02 models clearly depart from the observed trend.
The reason invoked for such discrepancy is the possible
spread in metallicity of galactic Cepheids, as recently shown by
Fry and Carney (1997).  However, the analysis of BCCM98
 implies that most
of the long period Cepheids observed are sub-metallic. We note
that two of the GFG98 Cepheids, T Mon ($\log \, P = 1.43$) and SV Vul
($\log P = 1.65$), are  metal-rich with  [Fe/H] = 0.09
and [Fe/H]=0.06 respectively, according to Fry and Carney (1997).
Moreover, if one assumes that the Galactocentric distance $R_{\rm gal}$ is
a crude indicator of metallicity (cf. GFG98), since the galactocentric
distance of Cepheids with  $\log \, P > 1.3$ in the sample of GFG98
is not systematically greater than  $R_{\rm gal}(\odot)$ (cf. 
Gieren et al. 1993), there is 
no reason to suspect that all these Cepheids have subsolar
metallicities (Fouqu\'e, priv. com.).
These arguments seem to exclude a possible selection effect in favor
of the detection of long period sub-metallic Cepheids and
weaken the argument of BCCM98.  

\subsubsection{LMC Cepheids}

Fig. 12 shows the comparison between the present standard models and
observations in period - magnitude diagrams for the LMC. The data in the $V$-band
are taken from GFG98, based on the sample of Tanvir (1997), from Laney and Stobie (1994, LS94) and from  the EROS1 sample (Sasselov et
al. 1997). The magnitudes provided by GFG98 and LS94 are
already dereddened. The EROS ($B_{\rm E}, R_{\rm E}$)
magnitudes are transformed in the Johnson $V$ magnitude according to 
Beaulieu et al. (1995), and for the reddening correction, 
we adopt $E(B-V)$=0.10
and $R_{\rm V}$ = 3.3, as in Baraffe et al. (1998). 
For the $J$- and $K$- bands, the data of GFG98 are used and
transformed in the CIT photometrical system. 
The distance modulus adopted in Fig. 12 is 18.50.

\begin{figure*}
\psfig{file=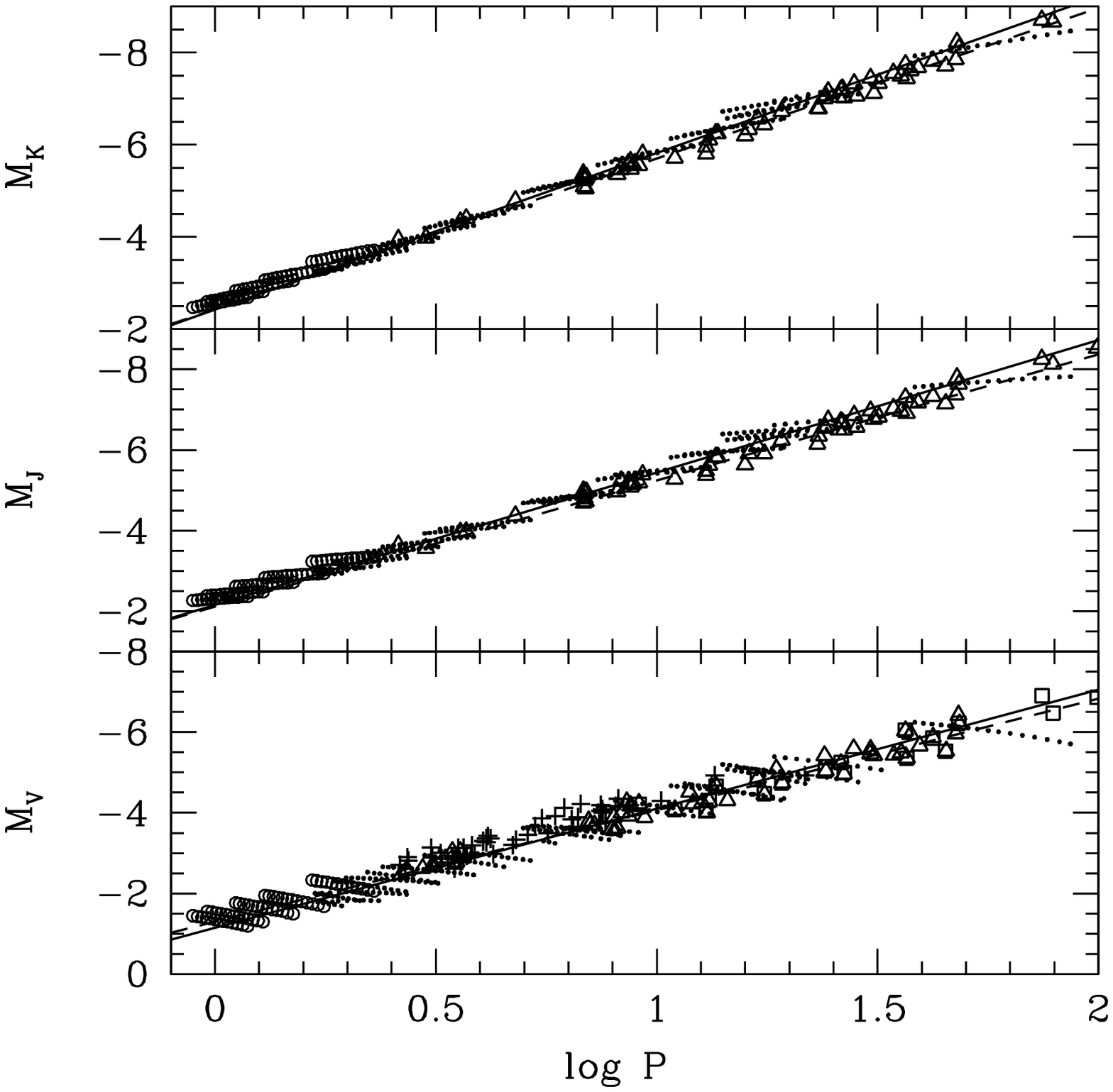,height=200mm,width=176mm}
\caption{Period - Magnitude diagrams in the $VJK$ bands for
 standard models with $Z$=0.01 and LMC observed Cepheids.
 The dots correspond to fundamental
unstable modes during core He burning phase from $\mmin$ = 3.875 $\msol$
to 12 $\msol$. 
The open circles correspond
to first crossing fundamental unstable modes for 
 3.875, 4, 4.25, 4.5, 5 $\msol$. Observations are taken from
GFG98 (open triangles), LS94 (open squares) and Sasselov et al. (1997)
 (+). The distance modulus is 18.50.
The curves have the same meaning as in Fig. 11.
}
\label{fig12}
\end{figure*}

The same excellent agreement is found between models and data
in {\it all} bandpasses.
Models with overshooting or with $l_{\rm mix}$ = 2
yield slightly different PL
relationships (\S. 3.3), 
but  we stress that it is impossible to 
discriminate between standard models
and models including overshooting when comparing to
bare data in period - magnitude diagrams.
Thus, comparison of
theoretical and observed period - magnitude diagrams and relationships cannot
be used to determine the amount of extra-mixing or the mixing
length parameter required in evolutionary models.  

A clear distinction between our standard models and models
with overshooting is however possible in terms of the minimum period observed.
As mentioned previously (cf. \S 2.3) and in Baraffe et al. (1998), standard
models for $Z$=0.01 predict fundamental pulsators during the blue loop
phase down to $\log \, P = 0.25$ (cf. Fig. 8), in contradiction with
 the observed end
of the EROS and MACHO PL distribution at $\log \, P \sim 0.4$. Standard
models with $Z$=0.008 predict even  lower period for the end
of the distribution, since $\mmin$ decreases. A slight increase
of $Y$  from 0.25 to 0.28 for the standard models with $Z$=0.01
slightly increases this period up to $\log \, P \sim 0.3$, 
 since it yields slightly brighter models 
and the same $\mmin$ (cf. \S 2.1 and Fig. 2). But this does not
solve the problem. Only models including overshooting
can yield better agreement, since they predict higher
$\mmin$ (cf. Fig. 9). This problem is further discussed with the
analysis of period histograms in \S 4.5.


\begin{table*}
\caption{Coefficients of the $\log \, P$ - Magnitude relationships
 (slope, zero-point)
for fundamental pulsators as
a function of metallicity for standard models.}
\begin{tabular}{lcccc}
\hline\noalign{\smallskip}
 & $M_V$ & $M_I$ & $M_J$ & $M_K$    \\
\noalign{\smallskip}
\hline\noalign{\smallskip}
$Z$=0.02  & (-2.905, -1.183) & (-3.102, -1.805) & (-3.256, -2.183) &
 (-3.367, -2.445) \\
$Z$=0.01  & (-2.951, -1.153) & (-3.140, -1.769) & (-3.286, -2.157) &
 (-3.395, -2.428) \\
$Z$=0.004 & (-2.939, -1.081) & (-3.124, -1.686) & (-3.262, -2.076) &
 (-3.369, -2.350) \\
\hline
\end{tabular}
\label{edge1cz43}
\end{table*}

\subsubsection{SMC Cepheids}

Comparison between SMC data and the present standard models
for $Z$=0.004 is shown in Fig. 13. Observations are from LS94,
which are already dereddened, and from Sasselov
et al. (1997) with an extinction coefficient $E(B-V)$=0.125 and
$R_{\rm V}$ = 3.3, as adopted in Baraffe et al. (1998). We adopt the
distance modulus $(m-M)_0 = 18.94$ of LS94.
Note that the SMC Cepheids observed by EROS are in the main
bar and the far arm of the SMC and are not likely to belong to the centroid
of the SMC (Beaulieu, priv. com.), resulting in a larger distance modulus difference with the 
LMC by about 0.15 mag compared to other studies (Caldwell \& Coulson 1986;
LS94). We however adopt the same distance modulus for 
both EROS and LS94 data for sake of simplicity.

\begin{figure*}
\psfig{file=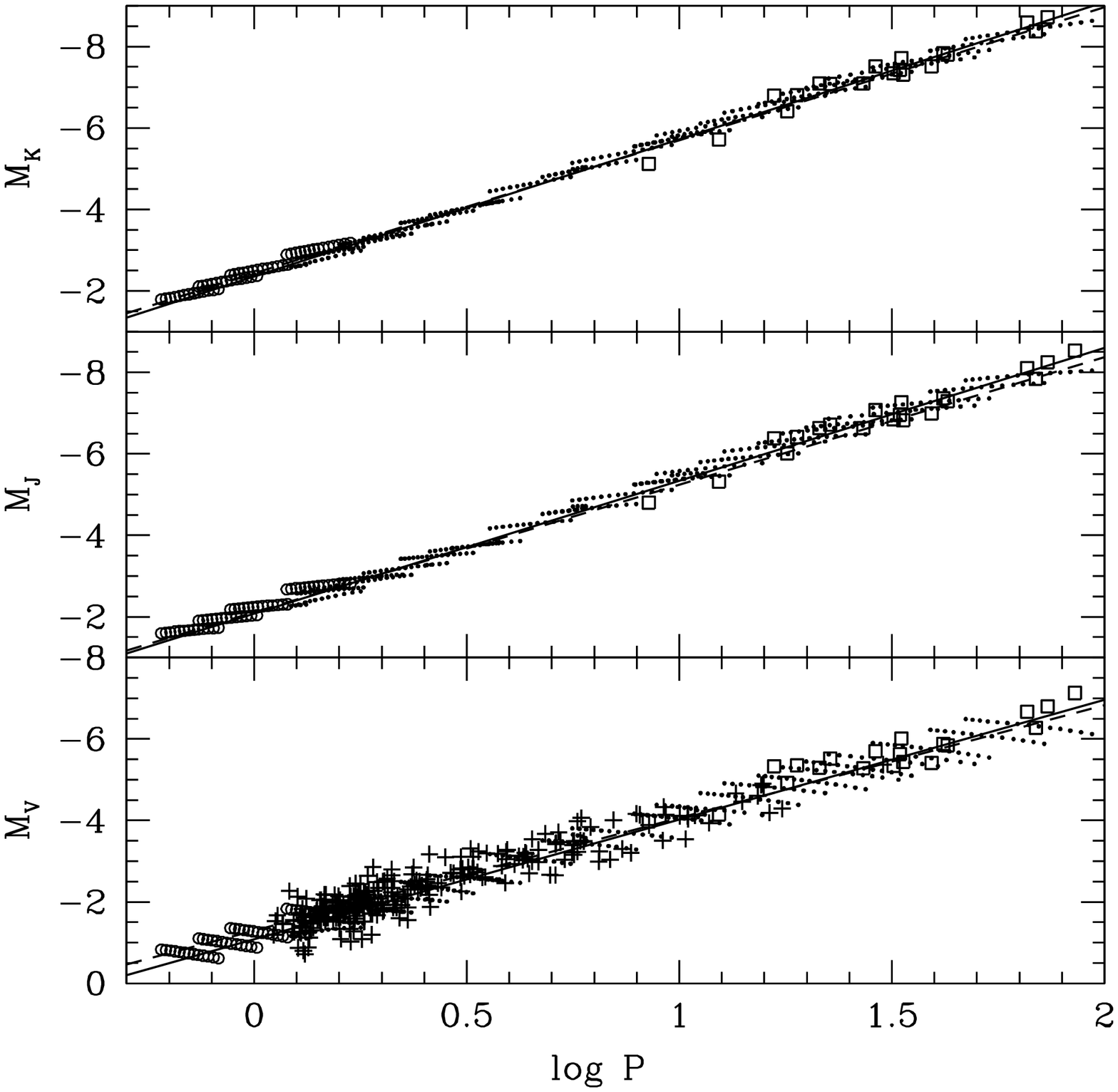,height=200mm,width=176mm}
\caption{Period - Magnitude diagrams in the $VJK$ bands for
 standard models with $Z$=0.004 and SMC observed Cepheids.
 The dots correspond to fundamental
unstable modes during core He burning phase from $\mmin$ = 3 $\msol$
to 12 $\msol$. 
The open circles correspond
to first crossing fundamental unstable modes for 
 3, 3.25, 3.5 and 4 $\msol$. Observations are taken from
 LS94 (open squares) and Sasselov et al. (1997)
 (+). The distance modulus is 18.94.
The curves have the same meaning as in Fig. 11.
}
\label{fig13}
\end{figure*}

The same agreement is reached as previously between data and models in Fig. 13,
as well as between the predicted period - magnitude relationships and
the observed one, although the GFG98 relationships are based
on galactic and LMC Cepheids. This already illustrates the small metallicity
effect predicted by the models on period - magnitude relationships,
as analysed in \S 5.

Regarding the observed minimum period, Baraffe et al. (1998) have
analysed  this point in detail , with the same EROS data.
 In this case, the standard models do a better job
and predict the correct position of the faintest Cepheid observed
(cf. Baraffe et al. 1998 and \S 2.3). The overshooting models
 predict a minimum period $\log \, P$ =  0.3 for models undergoing
a blue loop, which is excluded by observations (cf. Fig. 13
and Bauer et al. 1998). As for the LMC composition, and except for the
determination
of this minimum period, period - magnitude diagrams and
relationships cannot be used to discriminate standard
models from models with overshooting.

Our explanation by an evolutionary scenario 
of the observed  change
of slope near $P$ = 2 days for fundamental pulsators (cf. Baraffe et
al. 1998) has been questioned by Bauer et al. (1998), arguing that
such a change of slope should also
be observed for 1H Cepheids.
According
to Bauer et al. (1998), this change of slope is not apparent.
However, as mentioned in \S 3.4, this feature
is expected to be less pronounced for 1H than for fundamental pulsators
(cf. Fig. 10). Although  this could provide a nice explanation
for the observed Cepheids in the SMC, we remain
extremely cautious with such interpretation however, since we note 
several discrepancies between observed and predicted 1H pulsators. 
 Fig.14 shows the comparison of 1H
pulsators
with our standard models for $Z$=0.004 and $Z$=0.01 and the EROS SMC
and LMC data (Sasselov et al. 1997) respectively.
For $Z$=0.004, the models predict that 1H pulsators with $\log \, P \,
\simle 0$ belong to the first crossing phase, which concerns
 $\sim$ 1/10  of the observed 
1H pulsators. This is statistically difficult to explain
since comparison between the time-scales during
the first crossing and the blue loop phases predicts less than 1/100
of 1H pulsators in the first crossing. 
Regarding now the end of the distribution at high $P$, 
models predict unstable 1H modes up to $\log \, P \sim 1$, whereas 1H Cepheids
are not observed above $\log \, P \sim 0.5$, corresponding
to $\sim 5 \msol$ for the $Z$=0.004 standard models.
One possible explanation for such discrepancies is to
suppose that models in the first crossing phase, although
unstable in both F and 1H modes according to linear stability analysis,
favor 1H as the dominant mode. Similar reasons can be invoked
for $m \simgr 5 \msol$, for which this time F modes become
the dominant mode. Only non-linear calculations 
can test this
scenario. Note that if indeed first crossing models are preferably oscillating
in 1H mode, the higher number of this type of pulsators near
$\mmin$ provides an other explanation for the non detection
of the change of slope near $\mmin$, since first crossing models do 
not show any change of slope (cf. \S 3.4 and Fig. 10).

The other possible reason is a shortcoming due to
our neglect of the convection-pulsation coupling, yielding an erroneous
 IS. 
We indeed note that the
models do not reach the blue edge of the IS, for both the SMC
and the LMC (cf. Fig. 14). Models with overshooting or with $l_{\rm mix} =
2$
do not yield better agreement. 
Only a decrease of the distance
modulus of the SMC and the LMC by 0.3 mag can bring the predicted
and observed location of the 1H IS in agreement. However, with such smaller
distance moduli for both Clouds, the models do not reproduce
anymore the observed F pulsators in period - magnitude
diagrams. Since more data are available for F pulsators,
the models can be better tested against these observations, and we
rather believe distance moduli based on F pulsators. 
 Moreover,
we note that according to Yecko et al. (1998), the 1H IS seems to be more
sensitive to variations of parameters involved in the  convection-pulsation
coupling model than the F IS.  This point suggests
that  the neglect of the pulsation-convection
coupling in the present linear stability analysis affects 
1H modes more and yields an inaccurate IS, whereas fundamental modes
are less sensitive to such approximation, 
a suggestion supported by the general good agreement reached with F Cepheid observations.
This problem will be addressed in more details in a forthcoming paper
by analysing the uncertainties of the convective treatment
in the {\it pulsation} calculations. Period - magnitude
relationships for 1H will  be given in this forthcoming work.

\begin{figure*}
\psfig{file=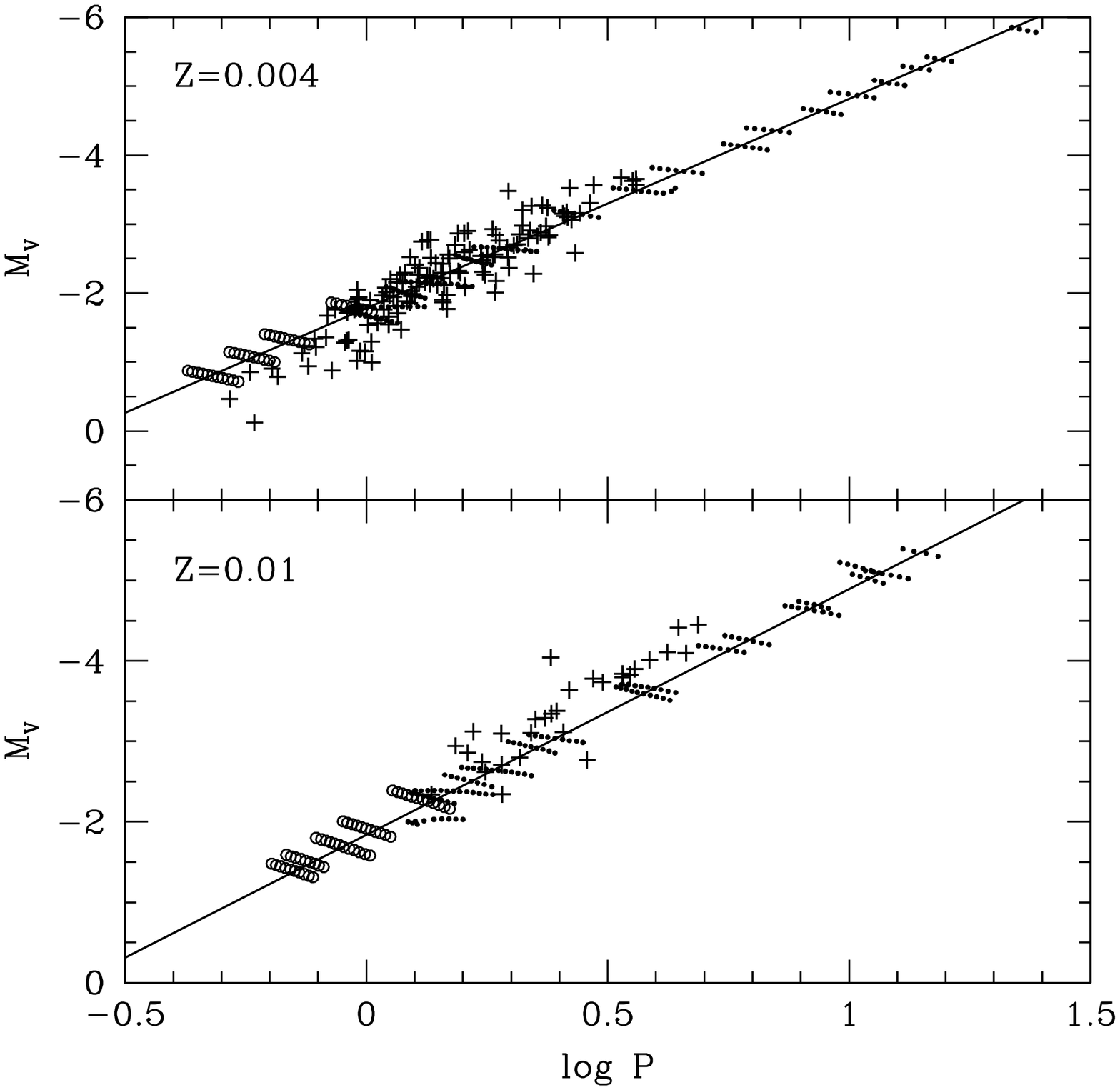,height=200mm,width=176mm}
\caption{Period - Magnitude diagram for 1H pulsators in the $V$ band for
 standard models with $Z$=0.01 and the EROS LMC data (lower
panel), and  $Z$=0.004 and the EROS SMC data (upper panel).
Symbols have the same meaning as in Fig. 12  for $Z$=0.01
and Fig. 13 for $Z$=0.004, but for
1H pulsators. 
The solid curves correspond to the $P$ - $\mv$ relationships derived from the models.
}
\label{fig14}
\end{figure*}

\subsection{ Period - Color relationships}

In this section, models are compared to galactic, LMC and SMC Cepheids
in period - color diagrams. As expected from the previous results, 
the uncertainties due to convection treatment in  stellar models
hardly affect the location of the IS in such diagrams.
We thus concentrate our analysis on standard models. 

Fig. 15 shows the comparison of galactic Cepheids with the $Z$=0.02
models, Fig. 16 corresponds to $Z$=0.01 models and LMC Cepheids and
Fig. 17
to $Z$=0.004 models and  SMC
Cepheids. The PC relationships (solid curves)
derived from our models
are displayed in each figures for the corresponding metallicity.
The coefficients are given in Tab. 4. The
PC relationships derived by LS94, 
based on the sample with reasonably
complete $BVJHK$ light curves, is shown in figures 15-17. These
observed
PC relationships 
 take into account the zero-point (ZP) shifts of LS94  for the Magellanic
Clouds.
First, we note the general good agreement for $(B-V)$ and $(V-K)$ colors
for the three metallicities displayed, although the predicted colors
seem to be too red compared to observations for $\log \, P \simle  1$. This trend
is more pronounced for the SMC data (cf. Fig. 17).
Regarding $(J-K)$ colors, the models
are bluer than observations by $\sim 0.05 - 0.1$ mag, this trend being
more pronounced for galactic and LMC data (cf. Fig. 15-16). 
We 
note as well that the positions of  objects in the theoretical IS
are fully consistent for $(B-V)$ and $(V-K$) colors, whereas the same objects
appear at a different location in the IS in the $P$ - $(J-K)$
diagram. This discrepancy appears also for SMC data (cf. Fig. 17),
although the models reproduce satisfactorily the data in $(J-K)$. The
possible sources of such discrepancies 
  are analysed in \S 5.

\begin{table}
\caption{Coefficients of the $\log \, P$ - color relationships
 (slope, zero-point)
for fundamental pulsators as
a function of metallicity for standard models.
}
\begin{tabular}{lccc}
\hline\noalign{\smallskip}
 & $B-V$ & $J-K$ & $V-K$    \\
\noalign{\smallskip}
\hline\noalign{\smallskip}
$Z$=0.02  & (0.246, 0.535) & (0.110, 0.261) & (0.462, 1.262) \\
$Z$=0.01  & (0.239, 0.497) & (0.108, 0.271) & (0.443, 1.275) \\
$Z$=0.004 & (0.231, 0.448) & (0.106, 0.274) & (0.429, 1.268) \\
\hline
\end{tabular}
\label{edge1cz43}
\end{table}

The present models give a better match to observations
than the Chiosi et al. (1993) models, which predict bluer
$(B-V)$ colors for the LMC and SMC Cepheids, and
the BCCM98 models, which  predict
significantly redder $(B-V)$ and $(V-K)$ colors for galactic Cepheids.
These two previous studies seem to overestimate the metallicity
effect on colors. Chiosi et al. (1993) attribute this discrepancy
to the ($B-V$, log $\te$) relations used in their analysis. 
In
the case of BCCM98, the discrepancies stem more
likely from the large metallicity effect they find on the
location of the IS, since an increase of $Z$ results in a
significant shift of their IS toward cooler $\te$. Whereas
their models reach good agreement in
period - magnitude and period-color diagrams with the SMC data, 
slight discrepancies appear already for the LMC data and
become larger in both diagrams for galactic Cepheids.
The large metallicity effect found by BCCM98 is
then questionable. The metallicity effects found
 on the present period - color
relationships are analysed and quantified in \S 5. 
 
\begin{figure*}
\psfig{file=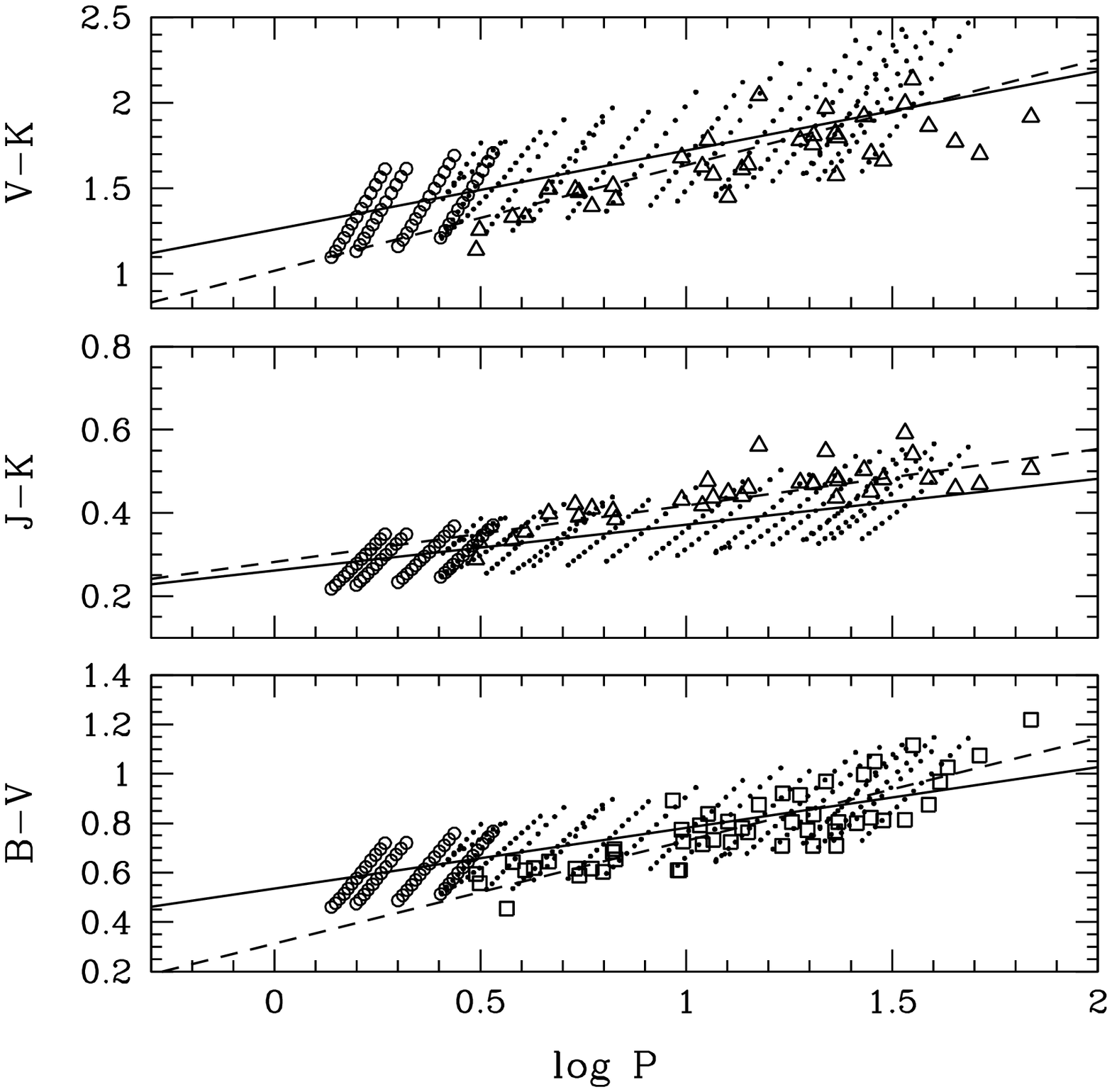,height=200mm,width=176mm}
\caption{Period - color diagrams for
solar metallicity standard models and galactic Cepheids.
 The dots correspond to fundamental
unstable modes during core He burning phase from $\mmin$ = 4.75 $\msol$
to 12 $\msol$. 
The open circles correspond
to first crossing fundamental unstable modes (see Fig. 11 for the
masses displayed). Observations  are from
GFG98 (open triangles) and LS94 (open squares). The dashed curves correspond to
the PC relationships given by LS94.
 The solid curves correspond
to the present PC relationships.
}
\label{fig15}
\end{figure*}

\begin{figure*}
\psfig{file=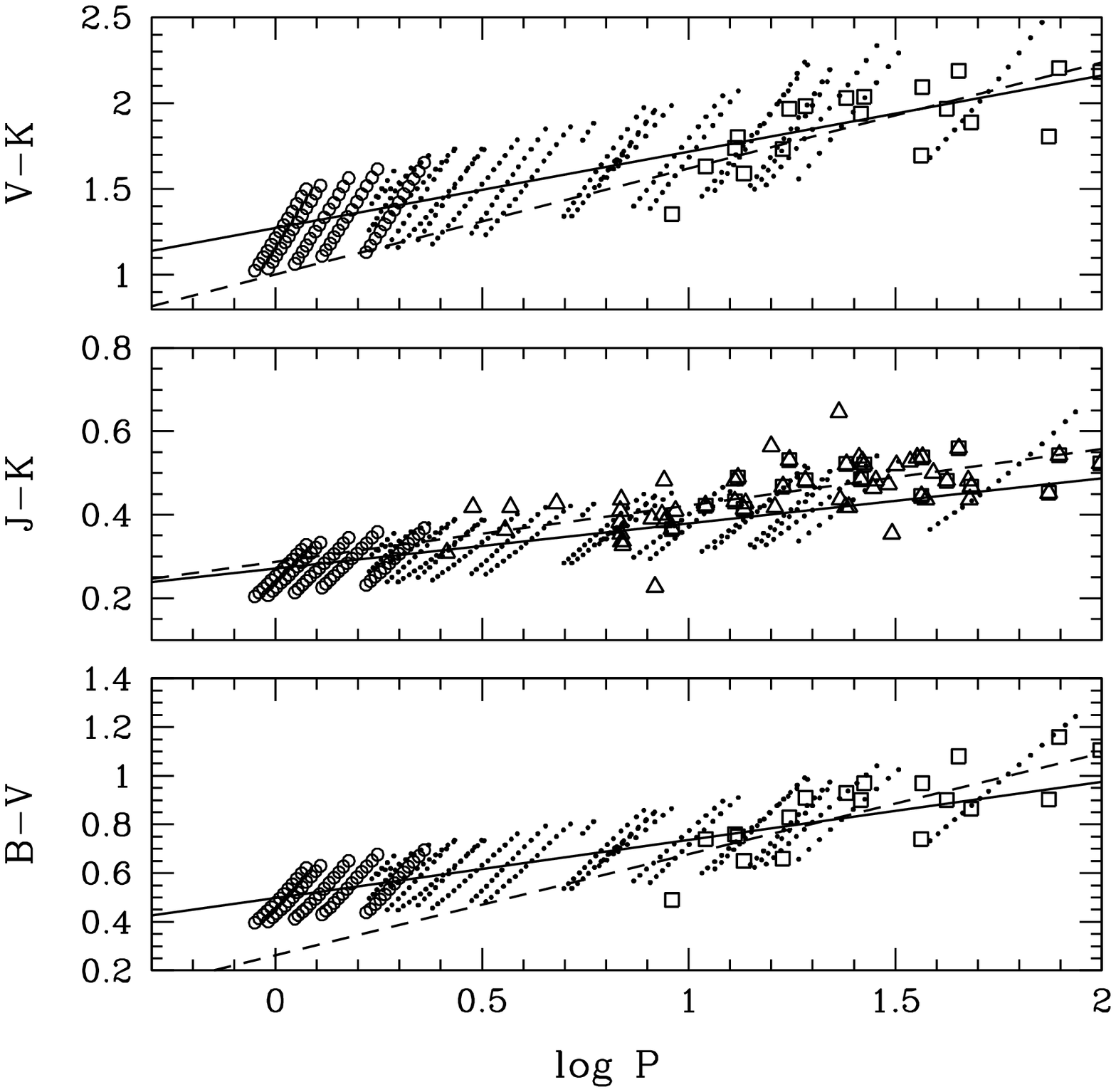,height=200mm,width=176mm}
\caption{Same as Fig. 15 for
$Z$=0.01 standard models and LMC Cepheids. The dots correspond to fundamental
unstable modes during core He burning phase from $\mmin$ = 3.875 $\msol$
to 12 $\msol$. 
The open circles correspond
to first crossing fundamental unstable modes 
(see Fig. 12 for the masses displayed).
 Observations are from
GFG98 (open triangles) and LS94 (open squares). The curves have the
same
meaning as in Fig. 15.
}
\label{fig16}
\end{figure*}

\begin{figure*}
\psfig{file=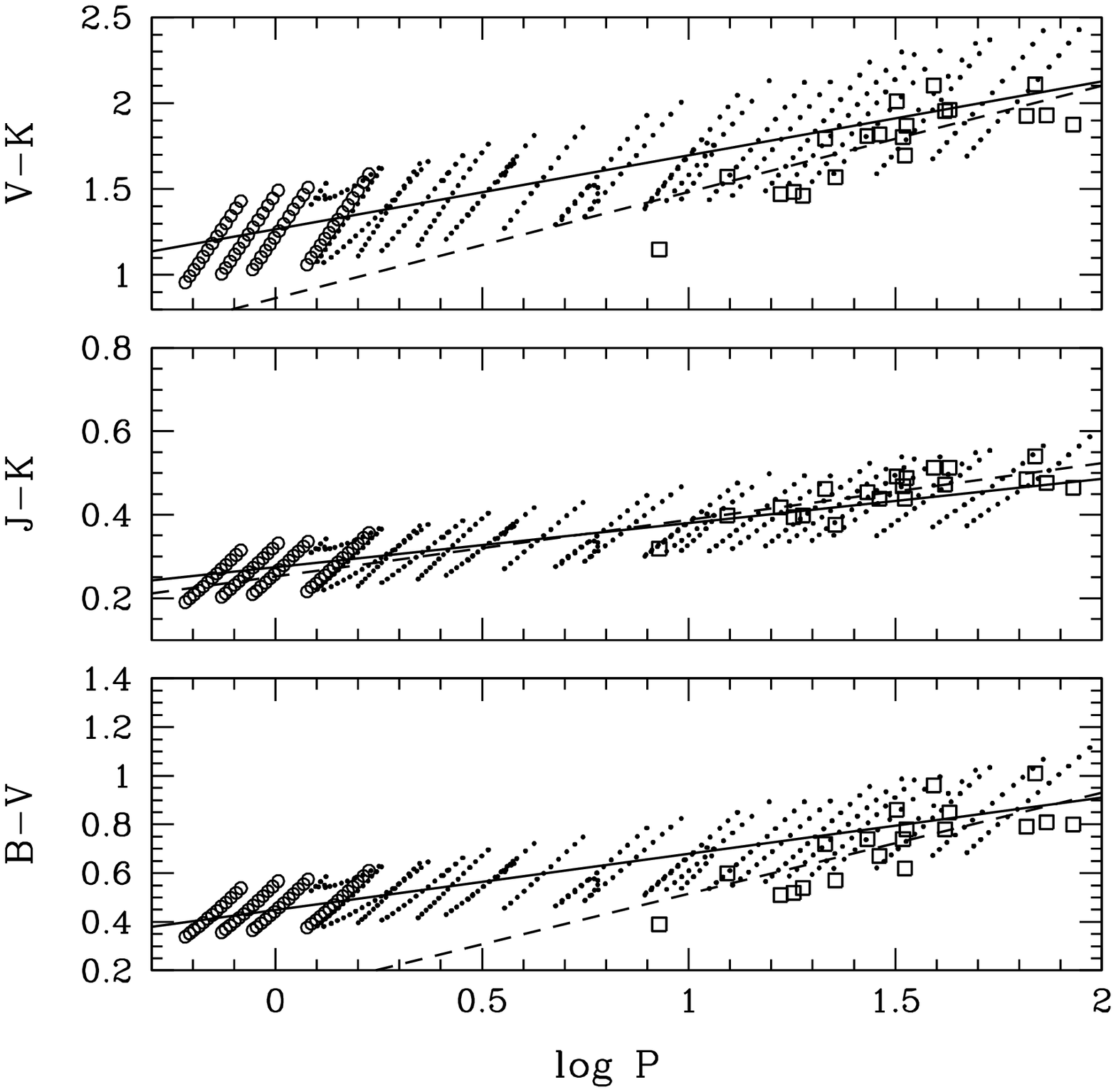,height=200mm,width=176mm}
\caption{Same as Fig. 15 for
$Z$=0.004 standard models and SMC Cepheids. The dots correspond to fundamental
unstable modes during core He burning  phase from $\mmin$ = 3 $\msol$
to 12 $\msol$. 
The open circles correspond
to first crossing fundamental unstable modes 
(see Fig. 13 for the masses displayed).
 Observations are from
 LS94 (open squares). The curves have the
same
meaning as in Fig. 15.
}
\label{fig17}
\end{figure*}

\subsection{Period - radius relationships}

Comparison between models and observations of GFG98 and Bersier et
al. (1997)
in a period - radius (PR) diagram
is performed in Fig. 18 for F and 1H pulsators. 
The general agreement for F pulsators is good, although
the models slightly overestimate  the radius for log $P < $ 1.
The comparison of the observed (solid line) 
and predicted PR relationships (dashed line)
illustrates better this discrepancy. For 1H pulsators, the observed
sample is unfortunately  too small to test the models.
Interestingly enough, we note
that the dispersion due to the width of the IS for
both F and 1H pulsators is almost
parallel to the relationship, resulting in a small intrinsic
dispersion due to the IS itself.  The coefficients of the PR relationships as
a function of $Z$ and convective treatment are given
for
fundamental pulsators in Tab. 5.

\begin{table}
\caption{Coefficients of the $\log \, P - \log \, R/R_\odot$ relationship (slope, zero-point)
 for fundamental pulsators as
a function of metallicity and convective treatment in the evolutionary
models.}
\begin{tabular}{lcc}
\hline\noalign{\smallskip}
($\alpha_{\rm mix}, d_{\rm ov}$) & (1.5,0) & (1.5,0.15)    \\
\noalign{\smallskip}
\hline\noalign{\smallskip}
$Z$=0.02 &(0.714, 1.143) & - \\  
$Z$=0.01 &(0.717, 1.142) &  (0.713, 1.123) \\  
$Z$=0.004 & (0.709, 1.129) & (0.734, 1.096) \\
\hline
\end{tabular}
\label{edge1cz43}
\end{table}

We find a small effect due to metallicity on the PR
relationship, with a maximum effect of 7\% on $R$ at a given
$P$.  
We find  as well small effects due to uncertainties
of convection treatment in the stellar models: models
with overshooting or with $l_{\rm mix}$=2 yield essentially the same
PR relationships, with a maximum effect of 7\% on $R$
at a given $P$. Note that the effect of uncertainties
due to convection in the 
 stellar models is of the same order as the effect of
$Z$. We therefore hardly expect this metallicity
effect to be observable. 

For 1H pulsators, the PR relationships for the $Z$=0.02
models shown in Fig. 18 is: 

\beq
\log {R \over R_{\odot}} = 0.723  \log P + 1.266.
\eeq

The effect of $Z$ on the 1H PR relationship is the same
as for fundamental pulsators.

Our results are rather similar to those of Bono et al. (1998b),
although they mention up
 to 9\% variation
of the radius due to metallicity effect. 
They find  the
same discrepancy between models and  the lowest period Cepheids
as mentioned  above. We urgently need more observational
data in order to determine whether this discrepancy stems 
from shortcomings  in the method used to
determine $R$ for low period Cepheids
(see Bono et al. (1998b) for discussion) or
from shortcomings in the models. 

\begin{figure*}
\psfig{file=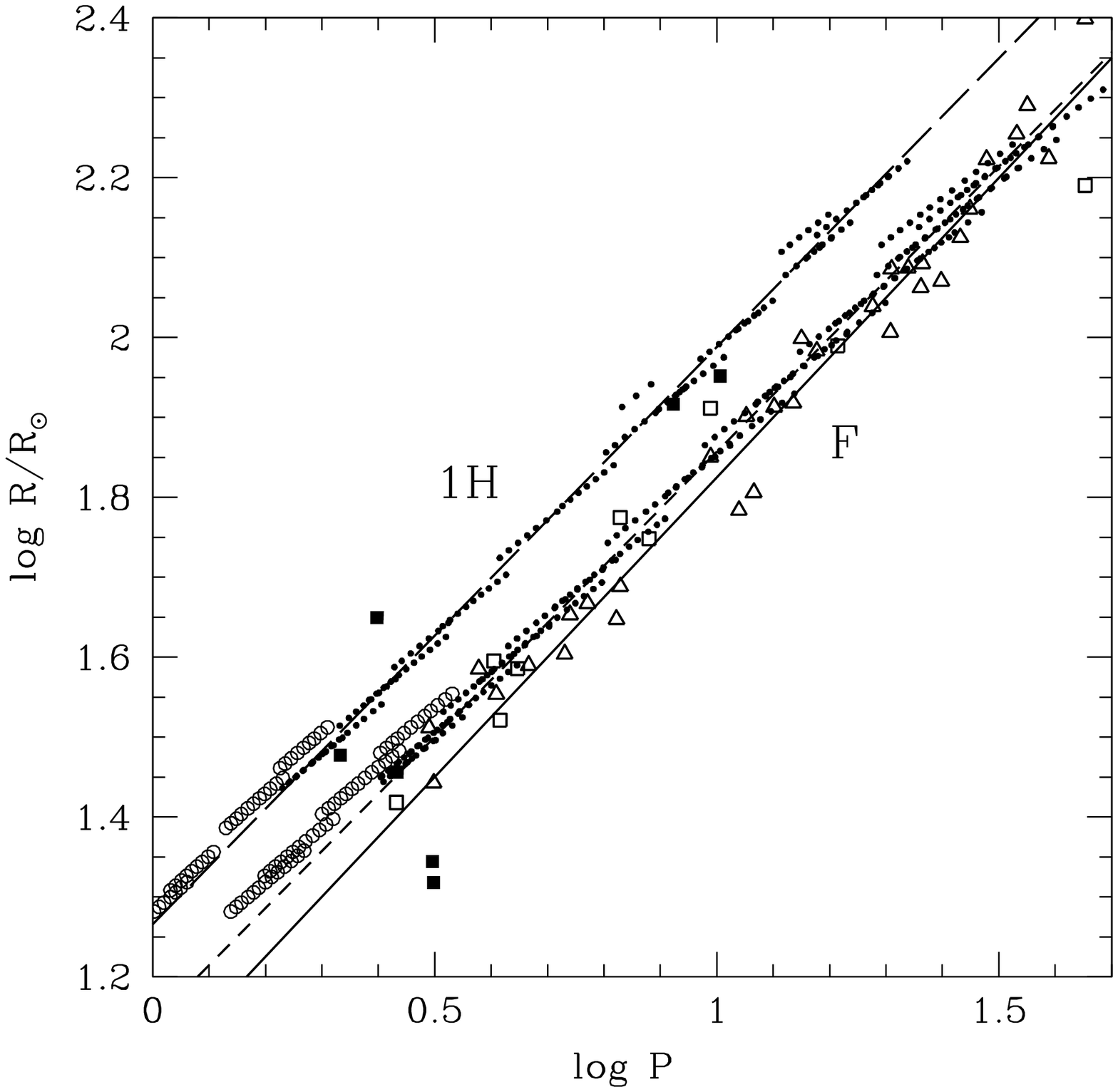,height=200mm,width=176mm}
\caption{Period - radius diagram for
solar metallicity standard models and galactic Cepheids,
for F (lower sequence) and 1H pulsators (upper sequence).
 Dots and open circles have the same meaning as in Fig. 11.
 Observations  are from
GFG98 (open triangles) and Bersier et al. (1997) for
classical (open squares) and {\it s-} Cepheids (full squares).
 The solid curves correspond to
the PR relationship given by GFG98.
 The dashed and long-dashed curves correspond
to the  present results for F and 1H pulsators respectively.
}
\label{fig18}
\end{figure*}

\subsection{Period-histograms}

The coupling between evolutionary and pulsation
calculations provides the opportunity to
construct period-histograms, taking into account a MF and evolutionary
timescales, as done for the statistical PL relationships derived
in Baraffe et al. (1998).
Since histograms give the number of Cepheids in a bin of period and  
are independent of photometrical, reddening or distance uncertainties,
they provide an interesting observational constraint.
The effect of  evolutionary times dominates that of 
the MF. In the following, a Salpeter MF is adopted with
an exponent -2.35. 
We have checked that a variation of the MF slope
from -2 to -4 does not alter the position of
the main peak and yields less than 8\% effect on the number of
objects in a period-bin.
 
For the SMC (cf. Fig. 19a), the models are compared to the 
period-histogram provided by the EROS-1  
collaboration (Sasselov et al. 1997).
The general shape predicted by the standard models is in good
agreement with observations. The bulk
of observed Cepheids at $\log \, P \sim 0.1 - 0.3$ is correctly reproduced
by the models and corresponds  
to models undergoing a blue loop near $\mmin$ = 3 $\msol$ (cf. Fig. 8).
We note however that
our calculations overestimate the number of these short-period Cepheids.
A similar distribution is obtained  for models with a different
 $l_{\rm mix}$.
As expected, the  histogram excludes calculations performed
with overshooting, which shifts the main peak to higher periods 
 by $\Delta \log \, P \simeq 0.2$.
  
\begin{figure*}
\psfig{file=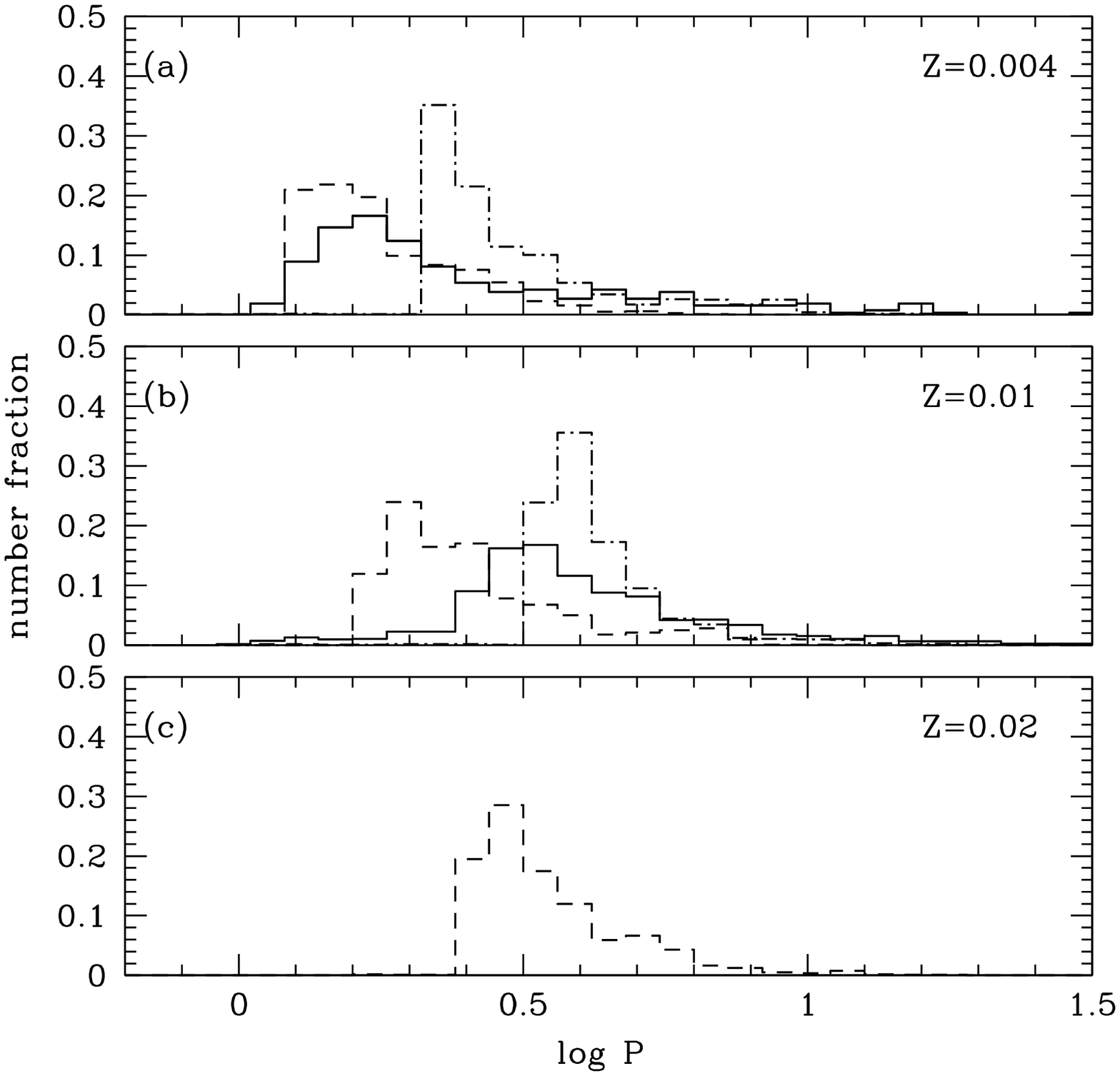,height=200mm,width=176mm}
\caption{{\bf (a)}: Period-histogram for the SMC. The solid line corresponds
 to the  EROS-1 observations (Sasselov et al. 1997).
The dashed line corresponds to the present standard models for $Z$=0.004, 
and the dash-dotted line to models with overshooting. 
{\bf (b):} same as (a) for LMC data and $Z$=0.01 models. The observed 
period-frequency distribution (solid line) is from MACHO (Alcock et al. 1998).
{\bf (c):} Same as (a) for $Z$=0.02 models.}
\label{histoSMC}
\end{figure*}

For LMC Cepheids (cf. Fig. 19b),  models are
compared to the fundamental period-frequency distribution provided recently
by Alcock et al. (1998). The standard models do 
not explain the observed histogram, as expected from the
discrepancy regarding the observed minimum period (cf. \S 4.2.2).
In this case, the theoretical histogram (dashed line) exhibits a peak
around
 $ \log P \sim 0.3$, instead
of the observed peak around $ \log P \sim 0.5$ in the MACHO
sample. Models including
overshooting could shift the peak toward the observed one. We note
that
at the time the present calculations with overshooting were performed, 
the only available sample was the EROS-1 LMC sample of data,
which shows a peak at $\log \, P \sim 0.6$. This motivated our
choice of the overshooting length $d_{\rm ov} = 0.15 H_{\rm P}$. The 
main peak of the MACHO LMC sample appears however at shorter
periods than the EROS data. 
In this case, a smaller $d_{\rm ov}$ than presently
adopted can provide
 better agreement with observations. The detailed
understanding of the observed period-histogram of LMC Cepheids 
 has been already analyzed by
Alcock et al. (1998). They note also difficulties to reproduce the 
observations, invoking a non constant star formation rate to explain
the long period tail after the maximum and anomalous Cepheids
resulting from binary coalescence to explain the short-period
Cepheids ($\log P \simle 0.4$). One of the arguments given in
Alcock et al. (1998) in favor of
such an exotic population of Cepheids concerns the double-mode
Cepheids, which according to Morgan and Welch (1997) can only 
be explained by non-standard ML relationships. However, Baraffe
et al. (1998) have shown that these double-mode Cepheids
can be explained naturally by first crossing models.
Such a suggestion is supported by statistical
arguments (cf. Baraffe et al. 1998). The presence of
short-period Cepheids below the main peak could
therefore be explained by first crossing models. Our main
concern is thus not
 the presence of these
short-period Cepheids, contrarily to Alcock et al. (1998), but
rather the failure of standard models to reproduce
the main peak of the LMC distribution, although they provide a good fit
to the SMC distribution (cf. Fig. 19a). This problem deserves further analysis
and definitely
requires the
confirmation of the minimum periods observed in the MC's
by similar surveys.

Finally, predictions for galactic models are displayed in Fig. 19c. 
The main peak is located at $\log \, P \sim 0.5$ for standard
models, with a sharp break at $\log \, P \sim 0.4$ (cf. Fig.8). 

\section{Discussion}

The general agreement obtained between present calculations
and observations in period - magnitude, -color  and -radius
diagrams gives us some confidence about the analysis 
of metallicity effects
based on the present models. We will first concentrate on 
 period - magnitude relationships. 
The effect of metallicity is more pronounced in the $B$ bandpass
in the range of interest $\te = 6500 {\rm K} - 4400$ K, due
to the dominant metallic line absorption, and it
decreases at longer wavelengths. 
For given gravity and $\te$ in the above-mentioned range,
a decrease of metallicity from $Z$=0.02 to $Z$=0.004 yields
a maximum increase by  $\sim$ 0.1 mag of the $B$-flux, and consequently
a decrease of the fluxes redward, of less than 0.05 mag in the $V$- and
$I$- bandpasses, 0.03 mag in  $J$ and 0.02 mag in $K$.

An inspection of the period - magnitude relationships
given in Tab. 3 show rather small effects of $Z$. 
A variation of $Z$ from 0.02
to 0.004 yields a difference in magnitude at a given $P$
 of less than 0.12 mag in $VIJK$ bandpasses
between $\log \, P = 0.5 - 2$.
Same quantitative effects of $Z$ are found with overshooting
models.

The comparison between predicted and observed (cf. GFG98)
PL relationships shows different slopes and zero-points, although
the data are well bracketed by the models (cf. Fig. 11-13). 
The resulting differences in magnitudes between our relationships and those
 of GFG98, for a given P, are however 
small, and amount to less than 0.15 mag in V, 0.25 mag in I, 0.35
mag in J and 0.25 mag in K. We note that these maximum discrepancies
appear for the highest periods observed. 
The largest discrepancy between observed and predicted relationships
appear in the $J$-band, although it
 can be reduced by $\sim$ 0.1 mag
by deriving a $P - M_{\rm J}$ relationship in the Carter system.
The models however seem to predict systematic higher $J$-fluxes, and
to a lesser extend $I$-fluxes, than GFG98 for a given $P$.

The present  effect of metallicity on the $P-M_V$ relationship is
of the same order as found by Chiosi et al. (1993), but their
relationships differ significantly from the present one and that
of GFG98 by more than 0.2 mag at $\log \, P > 1$, up
to $\sim 0.5$ mag at $\log \, P = 2$. 
As already mentioned, the present results significantly differ
from the results of BCCM98, who mention large metallicity effects,
up to $\delta M_{\rm V} \sim 0.77$ and $\delta M_K \sim 0.42 $
between the galactic and the SMC Cepheids. We note that their
$P-M_{\rm V}$ relationship for galactic Cepheids
yield discrepancies with the observed relationship of GFG98
of $\sim 0.4$ mag for $\log \, P > 1$ up to $\sim 1$ mag
at $\log \, P \sim 2$. As already mentioned, the possible contamination
of the sample by sub-metallic galactic Cepheids cannot be entirely 
responsible for such important discrepancies. 
The models of BCCM98 are based on non-linear calculations
accounting for the coupling between pulsation and convection.
According
to the systematic analysis of Yecko et al. (1998), the recipes used 
nowadays to describe the convection-pulsation coupling
 still require a certain number of free parameters, which
can significantly affect the location of the IS. In the work
of BCCM98, the sensitivity of the results to such uncertainties
is not clear, and should certainly be analysed. 
Moreover, as shown in \S 3.2 and Fig. 6, adopting
the same ML relationship for different Z, as done
in BCCM98, favors  
a shift of the IS location toward
cooler $\te$ as $Z$ increases. However, this effect cannot explain alone
the large shift found by BCCM98. 

Regarding now PC relationships, 
the effects of metallicity  are more detectable,
above all in $(B-V)$, as shown by models and observations in Fig. 15-17.
The comparison between the relationships given
in Tab. 4 shows that 
as $Z$ decreases from 0.02 to 0.004, $(B-V)$ gets bluer at a given $P$ 
by $\sim 0.1 - 0.12$ mag. 
A comparison of the blue and red edges of the IS
 between $Z$=0.02 and $Z$=0.004
shows the same shift for both edges toward the blue as Z decreases,
which reflects essentially the intrinsic effect of metallicity on this
color.
Similar effect is found by LS94, although slightly
larger, since LS94 mention a zero-point (ZP) shift between galactic and SMC
Cepheids of 0.215 mag.

For the $(J-K)$ colors, a variation of $Z$ from 0.02 to 0.004
 yields less than 0.01 mag
effect at a given $P$,  which is
consistent with the corresponding 0.033 mag effect quoted
by LS94. However, the models seem to predict $(J-K)$ (and $(H-K)$)
 colors  systematically bluer than observations by $\sim 0.05 - 0.1$ mag. 
The slight discrepancy mentioned previously for $M_{\rm J}$, and the
different positions of objects in the theoretical IS predicted
by $(J-K)$ colors compared to $(B-V$) or $(V-K)$ colors (cf. \S 4)
suggest a possible overestimate of the fluxes in the $IJH$ bands.

Finally, for (V-K) colors, the effect of metallicity is smaller
than 0.06 mag over the whole range of $P$ explored.
%
LS94 quotes a larger effect of $Z$ on this PC relationship,
with a shift of the ZP from  galactic to SMC Cepheids of 0.153.
BCCM98 find an extremely large effect, up to 0.65 mag at
$\log \, P = 2$, due to the shift of the IS toward cooler $\te$
as $Z$ increases. However, for {\it galactic Cepheids},
the BCCM98 PC relationships differ
by more than 0.2 mag compared to the LS94 relationship for $\log \, P > 1$,
 whereas we find no more than 0.1 mag difference for the same
range of period. For
the SMC composition, the $P - (V-K)$ relationship given by BCCM98
yields more than 0.2 mag difference in color compared to
the relation given by LS94, for $\log \, P > 1$,
whereas our results do not differ by more than 0.2 mag.  

  
Regarding the uncertainties of the present atmosphere models,
  we have neglected  the effects of
departures from local thermodynamic equilibrium, departures from
hydrostatic equilibrium on the atmospheric structure, and the effects
of varying the mixing length parameter. Since NLTE effects are
affecting the atomic line formation, we expect these to mainly affect
the synthetic spectra in the ultraviolet to visual spectral range
($U$, $B$, and $V$ bandpasses) where atomic transitions dominate the
spectral distribution of these stars. The propagation of hydrodynamic
chocs in the atmospheres may also affect the synthetic spectra by
modifying the thermal structure of the atmosphere. It is difficult to
estimate these effects at this point, but observation of such effects
in cooler Mira variable stars suggest a pronounced modification of the
spectra due to choc propagation in the near-infrared bandpasses ($J$,
$H$ and $K$). Convection, on the other hand, is limited to the
lowermost regions of the atmospheres ($\tau > 15$), and the synthetic
spectra are insensitive to changes of the mixing length parameter. 
All the possibilities mentioned above could be the source
of the observed color discrepancies quoted previously 
 and remain to be explored in subsequent works.

\section{Conclusion}

We have performed consistent stellar evolution
and pulsation calculations in order to determine evolutionary
effects on the properties of Cepheids as a function of metallicity.
We have investigated a range of metallicities representative of the Galaxy and
the MC populations.
The calculation of atmosphere models and synthetic spectra in
the present work
allows accurate comparisons with observations. We find {\it unprecedented}
levels of agreement with the most recent observations of Cepheids
in the Galaxy and the MC's in period - magnitude {\it and}
period - color diagrams. The good agreement with different
observational constraints  adds credibility
to the present models and to  the conclusions regarding the
metallicity effects.  
Our main results can be summarized as follow:

$\bullet$ A change of slope in the PL relationships is predicted
near the minimum mass $\mmin$ which undergoes a blue loop in
the instability strip during
core He burning. The critical period at which this change of slope
 occurs decreases with Z. For standard models, it takes
place at $\log \, P \sim 0.2$ for $Z$=0.004, $\log \, P \sim 0.35$
for $Z$=0.01 and  $\log \, P \sim 0.5$ for $Z$=0.02. These values are
affected by uncertainties due to convection
treatment in the stellar models, and increase when overshooting is taken
into account. However, the above mentioned $Z$ dependence 
is not affected by these uncertainties.   

$\bullet$ Taking into account the time spent in the IS for
each mass, the mean position is roughly in the middle
of the IS, except for masses near $\mmin$. This indicates
 that evolutionary timescales do not favor any particular position near
 the edges, in contradiction with the finding of
Fernie (1990). This result is independent of the convective treatment
used in the stellar models. A linear least-square fit to the models
above $\mmin$ is appropriate for
the determination of PL relationships.
   
$\bullet$ For a given Z, PL relationships derived on a mass range 
 or period range restricted to a region near $\mmin$ (or
to low periods) are much steeper than PL relationships
derived over the whole mass range considered in the present paper.
Thus, a {\it unique} linear least-square fit to the models down to
$\mmin$ is not appropriate. Since the period at which the
change of slope is expected depends on $Z$, 
 the comparison of PL relationships
derived on  Cepheids observed near this critical $P$
has to be taken with caution.

$\bullet$ The (linear) PL relationships derived in the present
work from $\mmin$ to 12 $\msol$ show a  weak dependence on
 metallicity, in agreement with a similar analysis performed by Saio and Gautschy (1998).

$\bullet$ A variation of $Z$ from 0.02 to 0.004
on Period - Magnitude relationships yields a difference in magnitude at a given P of
less than 0.12 mag in $VIJK$ bandpasses.   

$\bullet$ A decrease of $Z$ from 0.02
to 0.004 results in bluer $(B-V$) colors by 0.1 - 0.12 mag
at a given $P$, in agreement with observations. The effect
of $Z$ is found to be negligible for $(J-K)$.
 For $(V-K$) colors, we find a small
effect of $Z$ (less than 0.06 mag) at a given $P$.

 
$\bullet$ For first overtone pulsators, the models
predict  also a change of slope at the low end 
of the PL distribution, due to the reduction
of the He blue loop size. However, we expect
this feature to be less pronounced than for
F pulsators, due to the significant reduction
of the 1H instability strip width (by more than 400 K), compared
to the fundamental IS. 
According to the present
standard  models,
the observed 
SMC 1H pulsators with $\log \, P \simle 0$ should be in
the first crossing phase. Because of their relatively large observed 
number,
we suggest that first crossing models favor 1H as the dominant
pulsating mode. Non-linear calculations should be able to test this
hypothesis.

$\bullet$ The present standard models fail to reproduce the
minimum periods observed in the LMC by the EROS and MACHO collaborations,
and predict lower minimum periods. Models taking 
into account overshooting can better  account for these observations, but
they fail for the SMC Cepheids. Confirmation of the observed
minimum periods in the MC by the forthcoming MACHO and OGLE data
are urgently required to understand the source of this discrepancy.  

\medskip
The present conclusions regarding the metallicity effects remain
valid for different treatments of convection  used in the
stellar models. We certainly acknowledge that the present results may be
affected by  the neglect of the pulsation-convection coupling.
Since the present models yield
general good agreement with different observations for
{\it fundamental} pulsators, such approximation cannot be the source
of severe shortcomings. For 1H pulsators however, we quote
several discrepancies with observations, which may
indicate that the neglect of pulsation-convection coupling
is reasonable for the fundamental mode but not
for overtones. This problem will be addressed in a forthcoming work.

Finally, the present work suggests that for 
samples of Cepheids with $\log \, P \simgr 0.8$, metallicity
effects are negligible for galactic and MC environments,
and a universal PL relationship is certainly appropriate for
the derivation  of distance
moduli. This work suggests also that observed PL relationships
including low period Cepheids are to be taken with caution
and should be avoided to derive distance moduli, since evolutionary
effects break the linear form of this relationship at low P.  
The Z-effect suggested by Sasselov et al. (1997) in the low
$P$ regime is thus more likely the signature of these evolutionary
effects.

\medskip
Tables 6-8 are available by anonymous ftp:
\par
\hskip 1cm ftp ftp.ens-lyon.fr \par
\hskip 1cm username: anonymous \par
\hskip 1cm ftp $>$ cd /pub/users/CRAL/ibaraffe/cepheid \par
\hskip 1cm ftp $>$ get ABHA98\_models \par
\hskip 1cm ftp $>$ quit
\bigskip

Special requests for models can be addressed to Y. Alibert.

\begin{acknowledgements} We are grateful to our referee, A. Gautschy, for very
pertinent comments and for improving the manuscript. We thank
G. Chabrier
for numerous discussions and for reading
carefully the manuscript.
We thank P. Fouqu\'e for sending the filter transmission curves
of the Carter System and for many interesting discussions.
We thank also W. Glatzel for valuable discussions and U. Lee
for providing the original pulsation code. Part of
this work was performed during exchanges with 
the Universit\"ats-Sternwarte of G\"ottingen under
APAPE support (PROCOPE contract 97151). 
The calculations were performed on the T3E at Centre
d'Etudes Nucl\'eaires de Grenoble.
\end{acknowledgements}


\vfil\eject

\begin{table*}
\caption{Standard models with $(Z, Y)$ = (0.02, 0.28). The mass $m$ is in
$\msol$; $\te$ is the effective temperature
in K; $L$ corresponds to $\log \, L/L_\odot$; $t$ is the age in units
of $10^7$ yrs; $P_0$ is the fundamental mode period in days. The
$BVI$ absolute magnitudes are given in the Johnson-Cousins system 
and the $JHK$ magnitudes in the CIT system. For each mass, the blue (B)
and red (R) edges are given during all crossings of the IS.
For the first crossing, only models with $m \le 6
\msol$ are given.				       
}
\begin{tabular}{cccccccccccc}
\hline\noalign{\smallskip} 
edge & $m$ & $\te$ & $L$ & $t$ & $P_0$ & $M_{\rm B}$ & $M_{\rm V}$& $M_{\rm
I}$ & $M_{\rm J}$ & $M_{\rm H}$& $M_{\rm K}$\\
\noalign{\smallskip}
\hline\noalign{\smallskip}
B&   4.750 & 6304.7 &  2.710 &  9.8144 &  1.38 & -1.61 & -2.07 & -2.63 & -2.95 & -3.16 & -3.17\\
R&   4.750 & 5528.6 &  2.634 &  9.8184 &  1.86 & -1.06 & -1.78 & -2.55 & -3.04 & -3.36 & -3.39\\
R&   4.737 & 5331.8 &  2.848 & 10.9081 &  3.17 & -1.46 & -2.26 & -3.10 & -3.64 & -3.99 & -4.03\\
B&   4.737 & 5764.5 &  2.879 & 10.9573 &  2.57 & -1.80 & -2.43 & -3.13 & -3.57 & -3.84 & -3.87\\
R&   4.737 & 5332.4 &  2.905 & 11.5683 &  3.54 & -1.61 & -2.40 & -3.24 & -3.78 & -4.13 & -4.17\\
B&   5.000 & 6240.8 &  2.782 &  8.6764 &  1.58 & -1.77 & -2.25 & -2.82 & -3.15 & -3.37 & -3.38\\
R&   5.000 & 5526.1 &  2.713 &  8.6793 &  2.09 & -1.25 & -1.97 & -2.74 & -3.24 & -3.56 & -3.59\\
R&   4.985 & 5323.1 &  2.933 &  9.6454 &  3.64 & -1.67 & -2.47 & -3.31 & -3.85 & -4.21 & -4.24\\
B&   4.985 & 6089.2 &  2.986 &  9.6985 &  2.54 & -2.22 & -2.74 & -3.35 & -3.72 & -3.94 & -3.96\\
B&   4.985 & 6107.2 &  2.994 &  9.7986 &  2.55 & -2.25 & -2.76 & -3.36 & -3.73 & -3.95 & -3.97\\
R&   4.984 & 5253.6 &  3.010 & 10.4325 &  4.42 & -1.81 & -2.64 & -3.50 & -4.07 & -4.43 & -4.47\\
B&   5.500 & 6188.4 &  2.922 &  6.8681 &  2.00 & -2.10 & -2.59 & -3.17 & -3.52 & -3.74 & -3.75\\
R&   5.500 & 5423.5 &  2.851 &  6.8702 &  2.73 & -1.53 & -2.29 & -3.10 & -3.62 & -3.95 & -3.99\\
R&   5.477 & 5167.8 &  3.079 &  7.6340 &  5.04 & -1.92 & -2.79 & -3.68 & -4.27 & -4.65 & -4.69\\
B&   5.477 & 6042.6 &  3.136 &  7.6759 &  3.28 & -2.58 & -3.11 & -3.73 & -4.10 & -4.34 & -4.36\\
B&   5.476 & 6027.6 &  3.208 &  8.2280 &  3.79 & -2.75 & -3.29 & -3.91 & -4.29 & -4.52 & -4.54\\
R&   5.476 & 5147.6 &  3.182 &  8.3847 &  6.26 & -2.16 & -3.03 & -3.93 & -4.53 & -4.91 & -4.95\\
B&   6.000 & 6105.0 &  3.051 &  5.6114 &  2.53 & -2.39 & -2.91 & -3.51 & -3.87 & -4.10 & -4.12\\
R&   6.000 & 5403.3 &  2.987 &  5.6127 &  3.40 & -1.86 & -2.63 & -3.44 & -3.96 & -4.30 & -4.34\\
R&   5.965 & 5090.6 &  3.219 &  6.2371 &  6.62 & -2.21 & -3.11 & -4.03 & -4.64 & -5.03 & -5.08\\
B&   5.965 & 5928.8 &  3.268 &  6.2627 &  4.27 & -2.85 & -3.43 & -4.07 & -4.47 & -4.72 & -4.75\\
B&   5.963 & 5911.4 &  3.361 &  6.8159 &  5.17 & -3.08 & -3.65 & -4.31 & -4.71 & -4.96 & -4.98\\
R&   5.963 & 5116.4 &  3.331 &  6.8683 &  8.10 & -2.51 & -3.39 & -4.30 & -4.91 & -5.30 & -5.34\\
R&   6.932 & 4988.2 &  3.470 &  4.3770 & 10.55 & -2.75 & -3.70 & -4.65 & -5.30 & -5.71 & -5.76\\
B&   6.932 & 5852.1 &  3.503 &  4.3831 &  6.41 & -3.40 & -4.00 & -4.67 & -5.08 & -5.34 & -5.37\\
B&   6.927 & 5799.5 &  3.613 &  4.8696 &  8.20 & -3.65 & -4.26 & -4.95 & -5.37 & -5.64 & -5.66\\
R&   6.927 & 4930.9 &  3.579 &  4.8845 & 13.72 & -2.97 & -3.94 & -4.92 & -5.58 & -6.01 & -6.06\\
R&   7.886 & 4821.3 &  3.689 &  3.2799 & 16.98 & -3.14 & -4.17 & -5.19 & -5.89 & -6.34 & -6.40\\
B&   7.886 & 5764.9 &  3.722 &  3.2818 &  9.53 & -3.90 & -4.53 & -5.22 & -5.66 & -5.92 & -5.95\\
B&   7.876 & 5722.8 &  3.817 &  3.6870 & 11.80 & -4.11 & -4.75 & -5.46 & -5.91 & -6.18 & -6.21\\
R&   7.876 & 4859.0 &  3.781 &  3.6907 & 19.91 & -3.39 & -4.40 & -5.41 & -6.11 & -6.54 & -6.60\\
R&   8.835 & 4700.2 &  3.888 &  2.5871 & 25.80 & -3.50 & -4.59 & -5.67 & -6.42 & -6.88 & -6.95\\
B&   8.835 & 5642.7 &  3.918 &  2.5880 & 14.04 & -4.31 & -4.99 & -5.72 & -6.18 & -6.47 & -6.50\\
B&   8.819 & 5601.2 &  3.977 &  2.8848 & 16.23 & -4.43 & -5.12 & -5.87 & -6.34 & -6.63 & -6.67\\
R&   8.818 & 4660.5 &  3.935 &  2.8870 & 29.49 & -3.58 & -4.69 & -5.78 & -6.55 & -7.02 & -7.10\\
R&   9.822 & 4680.3 &  4.053 &  2.0969 & 34.18 & -3.89 & -4.99 & -6.07 & -6.83 & -7.30 & -7.37\\
B&   9.822 & 5543.9 &  4.080 &  2.0973 & 19.24 & -4.64 & -5.36 & -6.13 & -6.62 & -6.92 & -6.95\\
B&   9.791 & 5585.4 &  4.135 &  2.3821 & 20.99 & -4.81 & -5.51 & -6.26 & -6.74 & -7.03 & -7.06\\
R&   9.790 & 4589.3 &  4.089 &  2.3835 & 40.00 & -3.88 & -5.02 & -6.15 & -6.95 & -7.43 & -7.51\\
R&  10.883 & 4713.5 &  4.171 &  1.7332 & 39.41 & -4.20 & -5.29 & -6.36 & -7.11 & -7.57 & -7.64\\
B&  10.883 & 5521.4 &  4.193 &  1.7334 & 22.84 & -4.91 & -5.64 & -6.41 & -6.90 & -7.20 & -7.24\\
B&  10.826 & 5523.0 &  4.268 &  2.0405 & 26.77 & -5.09 & -5.82 & -6.59 & -7.08 & -7.39 & -7.42\\
R&  10.825 & 4610.3 &  4.224 &  2.0412 & 48.41 & -4.22 & -5.36 & -6.48 & -7.27 & -7.75 & -7.83\\
B&  11.940 & 5539.3 &  4.154 &  1.4633 & 19.59 & -4.83 & -5.55 & -6.32 & -6.80 & -7.10 & -7.14\\
R&  11.940 & 4643.6 &  4.098 &  1.4635 & 33.45 & -3.97 & -5.08 & -6.18 & -6.96 & -7.43 & -7.51\\
  \hline
  \end{tabular}
  \end{table*}

\vfil\eject

  \begin{table*}
  \caption{Same as in Table 6 for standard models with $(Z, Y)$ = (0.01,
  0.25). For the first crossing, only models with $m \le 5 \msol$ are given.
  }
  \begin{tabular}{cccccccccccc}
  \hline\noalign{\smallskip}
edge& $m$ & $\te$ & $L$ & $t$ & $P_0$ & $M_{\rm B}$ & $M_{\rm V}$& $M_{\rm
I}$ & $M_{\rm J}$ & $M_{\rm H}$& $M_{\rm K}$\\
  \noalign{\smallskip}
  \hline\noalign{\smallskip}
B&   3.875 & 6466.1 &  2.463 & 16.2120 &  0.89 & -1.05 & -1.45 & -1.96 & -2.27 & -2.46 & -2.47\\
R&   3.875 & 5703.9 &  2.394 & 16.2201 &  1.19 & -0.57 & -1.19 & -1.91 & -2.37 & -2.67 & -2.69\\
R&   3.870 & 5552.6 &  2.605 & 18.2885 &  1.94 & -1.02 & -1.69 & -2.45 & -2.94 & -3.27 & -3.30\\
B&   3.870 & 5819.8 &  2.625 & 18.3646 &  1.71 & -1.21 & -1.79 & -2.47 & -2.90 & -3.18 & -3.20\\
R&   3.870 & 5431.9 &  2.665 & 19.5117 &  2.35 & -1.10 & -1.82 & -2.61 & -3.14 & -3.48 & -3.51\\
B&   4.000 & 6441.0 &  2.507 & 15.0811 &  0.96 & -1.15 & -1.55 & -2.07 & -2.39 & -2.59 & -2.59\\
R&   4.000 & 5674.2 &  2.437 & 15.0882 &  1.28 & -0.67 & -1.30 & -2.02 & -2.48 & -2.79 & -2.82\\
R&   3.995 & 5528.2 &  2.658 & 17.0054 &  2.13 & -1.14 & -1.82 & -2.58 & -3.08 & -3.41 & -3.44\\
B&   3.995 & 6055.4 &  2.698 & 17.1486 &  1.69 & -1.50 & -2.00 & -2.61 & -3.00 & -3.24 & -3.26\\
R&   3.995 & 5380.8 &  2.734 & 18.3635 &  2.71 & -1.24 & -1.98 & -2.79 & -3.33 & -3.68 & -3.71\\
B&   4.250 & 6389.9 &  2.592 & 12.8715 &  1.11 & -1.35 & -1.77 & -2.30 & -2.62 & -2.82 & -2.83\\
R&   4.250 & 5610.9 &  2.521 & 12.8769 &  1.50 & -0.85 & -1.50 & -2.23 & -2.72 & -3.03 & -3.06\\
R&   4.244 & 5396.6 &  2.758 & 14.7441 &  2.70 & -1.31 & -2.04 & -2.85 & -3.38 & -3.73 & -3.77\\
B&   4.244 & 6207.4 &  2.815 & 14.8887 &  1.86 & -1.86 & -2.31 & -2.88 & -3.23 & -3.46 & -3.47\\
B&   4.244 & 6209.4 &  2.848 & 15.3763 &  1.98 & -1.94 & -2.39 & -2.96 & -3.31 & -3.54 & -3.55\\
R&   4.244 & 5385.5 &  2.842 & 15.9089 &  3.20 & -1.52 & -2.25 & -3.06 & -3.59 & -3.95 & -3.98\\
B&   4.500 & 6309.4 &  2.670 & 11.1956 &  1.30 & -1.52 & -1.95 & -2.50 & -2.84 & -3.05 & -3.06\\
R&   4.500 & 5539.0 &  2.600 & 11.1998 &  1.76 & -1.00 & -1.68 & -2.44 & -2.94 & -3.27 & -3.30\\
R&   4.493 & 5381.4 &  2.852 & 12.8880 &  3.15 & -1.54 & -2.28 & -3.08 & -3.62 & -3.98 & -4.01\\
B&   4.493 & 6138.6 &  2.905 & 13.0236 &  2.21 & -2.06 & -2.53 & -3.11 & -3.48 & -3.71 & -3.73\\
B&   4.493 & 6174.1 &  2.957 & 13.6604 &  2.40 & -2.20 & -2.66 & -3.24 & -3.60 & -3.83 & -3.84\\
R&   4.493 & 5315.9 &  2.933 & 13.9023 &  3.85 & -1.70 & -2.46 & -3.29 & -3.84 & -4.21 & -4.25\\
B&   5.000 & 6265.6 &  2.823 &  8.7200 &  1.66 & -1.89 & -2.33 & -2.89 & -3.23 & -3.45 & -3.46\\
R&   5.000 & 5487.8 &  2.757 &  8.7227 &  2.29 & -1.36 & -2.06 & -2.83 & -3.35 & -3.68 & -3.71\\
R&   4.990 & 5237.9 &  3.022 & 10.0765 &  4.51 & -1.87 & -2.66 & -3.52 & -4.09 & -4.47 & -4.51\\
B&   4.990 & 6071.5 &  3.077 & 10.2079 &  2.98 & -2.46 & -2.95 & -3.55 & -3.93 & -4.17 & -4.19\\
B&   4.990 & 6081.7 &  3.121 & 10.6372 &  3.23 & -2.58 & -3.06 & -3.66 & -4.04 & -4.28 & -4.29\\
R&   4.990 & 5221.5 &  3.088 & 10.7208 &  5.17 & -2.02 & -2.82 & -3.68 & -4.26 & -4.64 & -4.68\\
R&   5.977 & 5093.5 &  3.312 &  6.6542 &  7.79 & -2.48 & -3.33 & -4.24 & -4.86 & -5.27 & -5.31\\
B&   5.976 & 5910.4 &  3.356 &  6.7416 &  4.99 & -3.09 & -3.62 & -4.27 & -4.68 & -4.94 & -4.97\\
B&   5.976 & 5908.7 &  3.378 &  6.9163 &  5.22 & -3.14 & -3.68 & -4.33 & -4.74 & -5.00 & -5.02\\
R&   5.975 & 5112.1 &  3.350 &  6.9730 &  8.28 & -2.58 & -3.43 & -4.34 & -4.95 & -5.35 & -5.39\\
R&   6.954 & 5015.4 &  3.557 &  4.6978 & 12.09 & -3.02 & -3.92 & -4.85 & -5.49 & -5.92 & -5.96\\
B&   6.954 & 5818.6 &  3.579 &  4.7019 &  7.36 & -3.60 & -4.17 & -4.84 & -5.27 & -5.54 & -5.57\\
B&   6.952 & 5835.8 &  3.630 &  5.0668 &  8.07 & -3.73 & -4.30 & -4.96 & -5.39 & -5.66 & -5.68\\
R&   6.952 & 4985.1 &  3.589 &  5.0695 & 13.19 & -3.07 & -3.98 & -4.93 & -5.58 & -6.01 & -6.05\\
R&   7.921 & 4834.5 &  3.767 &  3.5386 & 19.40 & -3.37 & -4.36 & -5.36 & -6.07 & -6.53 & -6.58\\
B&   7.921 & 5731.7 &  3.788 &  3.5400 & 10.77 & -4.07 & -4.67 & -5.37 & -5.82 & -6.10 & -6.13\\
B&   7.918 & 5686.2 &  3.811 &  3.8251 & 11.58 & -4.10 & -4.72 & -5.43 & -5.89 & -6.18 & -6.21\\
R&   7.918 & 4818.1 &  3.751 &  3.8262 & 19.09 & -3.32 & -4.31 & -5.33 & -6.04 & -6.50 & -6.55\\
R&   8.880 & 4731.2 &  3.953 &  2.8145 & 28.55 & -3.72 & -4.76 & -5.82 & -6.56 & -7.04 & -7.10\\
B&   8.879 & 5640.2 &  3.978 &  2.8152 & 15.41 & -4.49 & -5.12 & -5.85 & -6.32 & -6.62 & -6.65\\
B&   8.873 & 5673.1 &  3.956 &  3.0381 & 14.45 & -4.45 & -5.08 & -5.79 & -6.25 & -6.54 & -6.58\\
R&   8.873 & 4932.3 &  3.899 &  3.0386 & 21.65 & -3.78 & -4.72 & -5.69 & -6.36 & -6.80 & -6.85\\
R&   9.930 & 4769.9 &  4.068 &  2.2168 & 32.22 & -4.04 & -5.07 & -6.10 & -6.83 & -7.30 & -7.36\\
B&   9.930 & 5590.1 &  4.090 &  2.2171 & 18.46 & -4.73 & -5.39 & -6.13 & -6.61 & -6.92 & -6.95\\
B&  10.979 & 5643.9 &  4.007 &  1.8114 & 14.11 & -4.57 & -5.20 & -5.92 & -6.39 & -6.69 & -6.72\\
R&  10.979 & 4861.9 &  3.954 &  1.8115 & 21.96 & -3.86 & -4.83 & -5.83 & -6.53 & -6.98 & -7.03\\
B&  11.874 & 5429.4 &  4.454 &  1.8127 & 38.40 & -5.51 & -6.24 & -7.04 & -7.56 & -7.89 & -7.93\\
R&  11.868 & 4392.1 &  4.430 &  1.8160 & 86.39 & -4.47 & -5.71 & -6.94 & -7.82 & -8.37 & -8.47\\
  \hline
    \end{tabular}
      \end{table*}

\begin{table*}
\caption{Same as in Table 6 for standard models with $(Z, Y)$ = (0.004,
0.25).
For the first crossing, only models with $m \le 4 \msol$ are given
}
\begin{tabular}{cccccccccccc}
\hline\noalign{\smallskip}
edge& $m$ & $\te$ & $L$ & $t$ & $P_0$ & $M_{\rm B}$ & $M_{\rm V}$& $M_{\rm
I}$ & $M_{\rm J}$ & $M_{\rm H}$& $M_{\rm K}$\\
\noalign{\smallskip}
\hline\noalign{\smallskip}
B&   3.000 & 6631.0 &  2.225 & 26.0600 &  0.60 & -0.49 & -0.83 & -1.31 & -1.60 & -1.79 & -1.79\\
R&   3.000 & 5837.0 &  2.166 & 26.0772 &  0.82 & -0.08 & -0.62 & -1.29 & -1.73 & -2.03 & -2.05\\
R&   2.998 & 5679.9 &  2.370 & 29.3156 &  1.32 & -0.52 & -1.11 & -1.83 & -2.30 & -2.62 & -2.64\\
B&   2.998 & 5858.8 &  2.384 & 29.4170 &  1.22 & -0.64 & -1.17 & -1.83 & -2.27 & -2.56 & -2.58\\
R&   2.998 & 5550.5 &  2.483 & 32.0947 &  1.78 & -0.74 & -1.37 & -2.12 & -2.63 & -2.97 & -2.99\\
B&   3.250 & 6534.5 &  2.335 & 21.4880 &  0.74 & -0.74 & -1.10 & -1.60 & -1.90 & -2.10 & -2.11\\
R&   3.250 & 5741.4 &  2.273 & 21.4996 &  1.02 & -0.31 & -0.88 & -1.58 & -2.04 & -2.35 & -2.37\\
R&   3.248 & 5556.0 &  2.520 & 24.7568 &  1.81 & -0.84 & -1.46 & -2.21 & -2.72 & -3.06 & -3.08\\
B&   3.248 & 6389.0 &  2.582 & 25.2131 &  1.26 & -1.33 & -1.71 & -2.24 & -2.57 & -2.79 & -2.79\\
B&   3.248 & 6396.7 &  2.603 & 25.6821 &  1.31 & -1.39 & -1.77 & -2.29 & -2.62 & -2.84 & -2.84\\
R&   3.248 & 5496.7 &  2.637 & 27.0614 &  2.34 & -1.10 & -1.74 & -2.51 & -3.03 & -3.37 & -3.40\\
B&   3.500 & 6485.7 &  2.438 & 17.9419 &  0.88 & -0.99 & -1.36 & -1.87 & -2.18 & -2.38 & -2.39\\
R&   3.500 & 5716.3 &  2.379 & 17.9499 &  1.20 & -0.56 & -1.14 & -1.84 & -2.31 & -2.62 & -2.64\\
R&   3.498 & 5514.7 &  2.643 & 20.7979 &  2.23 & -1.12 & -1.76 & -2.53 & -3.04 & -3.38 & -3.41\\
B&   3.498 & 6326.6 &  2.711 & 21.3714 &  1.59 & -1.64 & -2.03 & -2.57 & -2.91 & -3.13 & -3.14\\
B&   3.498 & 6269.9 &  2.765 & 22.3496 &  1.81 & -1.75 & -2.16 & -2.72 & -3.07 & -3.29 & -3.30\\
R&   3.498 & 5449.5 &  2.747 & 22.5792 &  2.84 & -1.35 & -2.01 & -2.79 & -3.32 & -3.67 & -3.70\\
B&   4.000 & 6424.4 &  2.628 & 13.0369 &  1.19 & -1.46 & -1.83 & -2.35 & -2.68 & -2.89 & -2.89\\
R&   4.000 & 5601.6 &  2.567 & 13.0418 &  1.69 & -0.98 & -1.59 & -2.33 & -2.82 & -3.15 & -3.18\\
R&   3.998 & 5362.6 &  2.835 & 14.9299 &  3.27 & -1.51 & -2.21 & -3.02 & -3.57 & -3.94 & -3.97\\
B&   3.998 & 6212.6 &  2.899 & 15.3241 &  2.21 & -2.08 & -2.49 & -3.06 & -3.42 & -3.65 & -3.67\\
B&   3.998 & 6151.7 &  2.964 & 16.2167 &  2.59 & -2.22 & -2.65 & -3.23 & -3.60 & -3.84 & -3.86\\
R&   3.998 & 5297.1 &  2.944 & 16.3816 &  4.23 & -1.74 & -2.47 & -3.30 & -3.86 & -4.24 & -4.28\\
R&   4.997 & 5237.8 &  3.140 &  8.5581 &  5.55 & -2.19 & -2.94 & -3.79 & -4.37 & -4.76 & -4.80\\
B&   4.997 & 6043.7 &  3.176 &  8.6299 &  3.58 & -2.71 & -3.17 & -3.78 & -4.17 & -4.43 & -4.44\\
B&   4.997 & 6003.3 &  3.313 &  9.6985 &  4.77 & -3.04 & -3.51 & -4.12 & -4.52 & -4.78 & -4.80\\
R&   4.997 & 5156.7 &  3.290 &  9.7818 &  7.92 & -2.50 & -3.29 & -4.17 & -4.77 & -5.18 & -5.22\\
R&   5.995 & 5069.4 &  3.417 &  5.6779 &  9.59 & -2.75 & -3.58 & -4.49 & -5.12 & -5.54 & -5.58\\
B&   5.995 & 5932.8 &  3.436 &  5.6864 &  5.60 & -3.32 & -3.80 & -4.44 & -4.86 & -5.12 & -5.14\\
B&   5.994 & 5832.6 &  3.581 &  6.5151 &  7.88 & -3.63 & -4.15 & -4.81 & -5.25 & -5.53 & -5.55\\
R&   5.993 & 5024.9 &  3.556 &  6.5638 & 13.12 & -3.05 & -3.90 & -4.83 & -5.47 & -5.91 & -5.95\\
R&   6.988 & 4937.2 &  3.664 &  4.1333 & 15.67 & -3.24 & -4.14 & -5.10 & -5.77 & -6.22 & -6.27\\
B&   6.988 & 5787.3 &  3.677 &  4.1355 &  8.84 & -3.85 & -4.39 & -5.06 & -5.51 & -5.79 & -5.82\\
B&   6.986 & 5777.6 &  3.795 &  4.7436 & 11.22 & -4.14 & -4.67 & -5.35 & -5.80 & -6.09 & -6.11\\
R&   6.986 & 4947.7 &  3.763 &  4.7451 & 19.06 & -3.49 & -4.38 & -5.34 & -6.01 & -6.46 & -6.50\\
R&   7.977 & 4945.3 &  3.881 &  3.2176 & 22.02 & -3.78 & -4.67 & -5.64 & -6.30 & -6.75 & -6.80\\
B&   7.977 & 5740.7 &  3.894 &  3.2185 & 12.77 & -4.36 & -4.91 & -5.60 & -6.06 & -6.35 & -6.38\\
B&   7.972 & 5701.6 &  3.972 &  3.6736 & 15.35 & -4.53 & -5.10 & -5.80 & -6.26 & -6.56 & -6.59\\
R&   7.972 & 4829.4 &  3.938 &  3.6745 & 27.44 & -3.81 & -4.77 & -5.78 & -6.48 & -6.95 & -7.00\\
R&   8.972 & 4772.5 &  4.043 &  2.5642 & 32.77 & -4.01 & -5.00 & -6.03 & -6.76 & -7.24 & -7.30\\
B&   8.972 & 5641.5 &  4.059 &  2.5647 & 17.52 & -4.72 & -5.31 & -6.03 & -6.50 & -6.81 & -6.84\\
B&   8.964 & 5625.3 &  4.117 &  2.9297 & 19.90 & -4.85 & -5.44 & -6.17 & -6.65 & -6.96 & -6.99\\
R&   8.964 & 4786.1 &  4.081 &  2.9303 & 35.16 & -4.11 & -5.10 & -6.13 & -6.85 & -7.33 & -7.38\\
R&   9.987 & 4767.9 &  4.181 &  2.0752 & 40.58 & -4.33 & -5.33 & -6.37 & -7.10 & -7.58 & -7.64\\
B&   9.987 & 5564.1 &  4.193 &  2.0755 & 22.46 & -5.00 & -5.62 & -6.37 & -6.86 & -7.18 & -7.21\\
B&   9.962 & 5562.2 &  4.307 &  2.4855 & 28.54 & -5.28 & -5.90 & -6.64 & -7.14 & -7.46 & -7.49\\
R&   9.962 & 4713.1 &  4.283 &  2.4864 & 53.56 & -4.51 & -5.54 & -6.61 & -7.36 & -7.86 & -7.91\\
B&  10.938 & 5450.5 &  4.450 &  2.1391 & 38.98 & -5.55 & -6.22 & -7.01 & -7.53 & -7.86 & -7.90\\
R&  10.937 & 4669.6 &  4.433 &  2.1403 & 72.15 & -4.81 & -5.88 & -6.97 & -7.74 & -8.24 & -8.30\\
B&  11.912 & 5428.8 &  4.562 &  1.8561 & 47.16 & -5.80 & -6.49 & -7.28 & -7.81 & -8.14 & -8.18\\
R&  11.909 & 4596.2 &  4.548 &  1.8585 & 93.53 & -5.01 & -6.12 & -7.25 & -8.04 & -8.57 & -8.63\\
  \hline
    \end{tabular}
      \end{table*}

\end{document}